\documentclass[aps,
nofootinbib,
superscriptaddress,
showpacs,twocolumn,
pra,longbibliography]{revtex4-1}
\usepackage{amsmath}
\usepackage{easybmat}
\usepackage{amsfonts}  
\usepackage{enumitem}
\usepackage{graphicx}
\usepackage{amsthm}
\usepackage{subfigure}
\usepackage{array}
\usepackage{hyperref}
\usepackage[usenames]{color}
\usepackage{centernot}
\usepackage{moreverb}
\usepackage{rotating}

\usepackage{braket}

\newcommand{\lp}{\left ( }
\newcommand{\rp}{\right ) }

\newcommand{\hc}{\text{H.c.}}

\newcommand{\beq}{\begin{eqnarray*}}
\newcommand{\eeq}{\end{eqnarray*}}

\newcommand{\be}{\begin{eqnarray}}
\newcommand{\ee}{\end{eqnarray}}

\newcommand{\der}[2]{\frac{d #1}{d #2}}

\newcommand{\secder}[2]{\frac{d^2 #1}{d #2^2}}

\newcommand{\diff}[2]{\frac{\partial #1}{\partial #2}}
\newcommand{\secdiff}[2]{\frac{\partial^2 #1}{\partial #2^2}}

\newcommand{\mc}{\mathcal}

\newcommand{\rvdw}{R_{\text{vdW}}}
\newcommand{\lho}{l_{\text{ho}}}

\newcommand{\assignment}[1]{}

\def\lsim{\mathrel{\rlap{\lower4pt\hbox{\hskip1pt$\sim$}}
    \raise1pt\hbox{$<$}}}                % less than or approx.

\def\gsim{\mathrel{\rlap{\lower4pt\hbox{\hskip1pt$\sim$}}
    \raise1pt\hbox{$>$}}}                % greater than or approx.

\begin{document}

\title{Microscopic derivation of multi-channel Hubbard models for ultracold nonreactive molecules in an optical lattice} 
\author{Michael L. Wall} 
\thanks{Present address: The Johns Hopkins University Applied Physics Laboratory, Laurel, MD 20723, USA}
\affiliation{JILA, NIST and University of Colorado, Boulder, Colorado 80309, USA}
\author{Nirav P.~Mehta} 
\affiliation{Department of Physics and Astronomy, Trinity University, San Antonio, Texas 78212, USA}
\author{Rick Mukherjee}  
\affiliation{Department of Physics and Astronomy, Rice University, Houston, Texas 77005, USA}
\affiliation{Rice Center for Quantum Materials, Rice University, Houston, Texas 77005, USA}
\author{Shah Saad Alam}  
\affiliation{Department of Physics and Astronomy, Rice University, Houston, Texas 77005, USA}
\affiliation{Rice Center for Quantum Materials, Rice University, Houston, Texas 77005, USA}
\author{Kaden R.~A. Hazzard} \email{kaden.hazzard@gmail.com}
\affiliation{Department of Physics and Astronomy, Rice University, Houston, Texas 77005, USA}
\affiliation{Rice Center for Quantum Materials, Rice University, Houston, Texas 77005, USA}

\begin{abstract}
Recent experimental advances in the cooling and manipulation of bialkali dimer molecules have enabled the production of gases of ultracold molecules that are not chemically reactive.  It has been presumed in the literature that in the absence of an electric field the low-energy scattering of such nonreactive molecules (NRMs) will be similar to atoms, in which a single $s$-wave scattering length governs the collisional physics.  However, in Ref.~\cite{docaj:ultracold_2016}, it was argued that the short-range collisional physics of NRMs is much more complex than for atoms, and that this leads to a many-body description in terms of a multi-channel Hubbard model.  In this work, we show that this multi-channel Hubbard model description of NRMs in an optical lattice is robust against the approximations employed in Ref.~\cite{docaj:ultracold_2016} to estimate its parameters.  We do so via an exact, albeit formal, derivation of a multi-channel resonance model for two NRMs from an ab initio description of the molecules in terms of their constituent atoms.  We discuss the regularization of this two-body multi-channel resonance model in the presence of a harmonic trap, and how its solutions form the basis for the many-body model of Ref.~\cite{docaj:ultracold_2016}.  We also generalize the derivation of the effective lattice model to include multiple internal states (e.g., rotational or hyperfine).  We end with an outlook to future research.
\end{abstract}

\maketitle

\section{Introduction \label{sec:introduction}}

\begin{figure}[t]
\includegraphics[width=0.9\columnwidth]{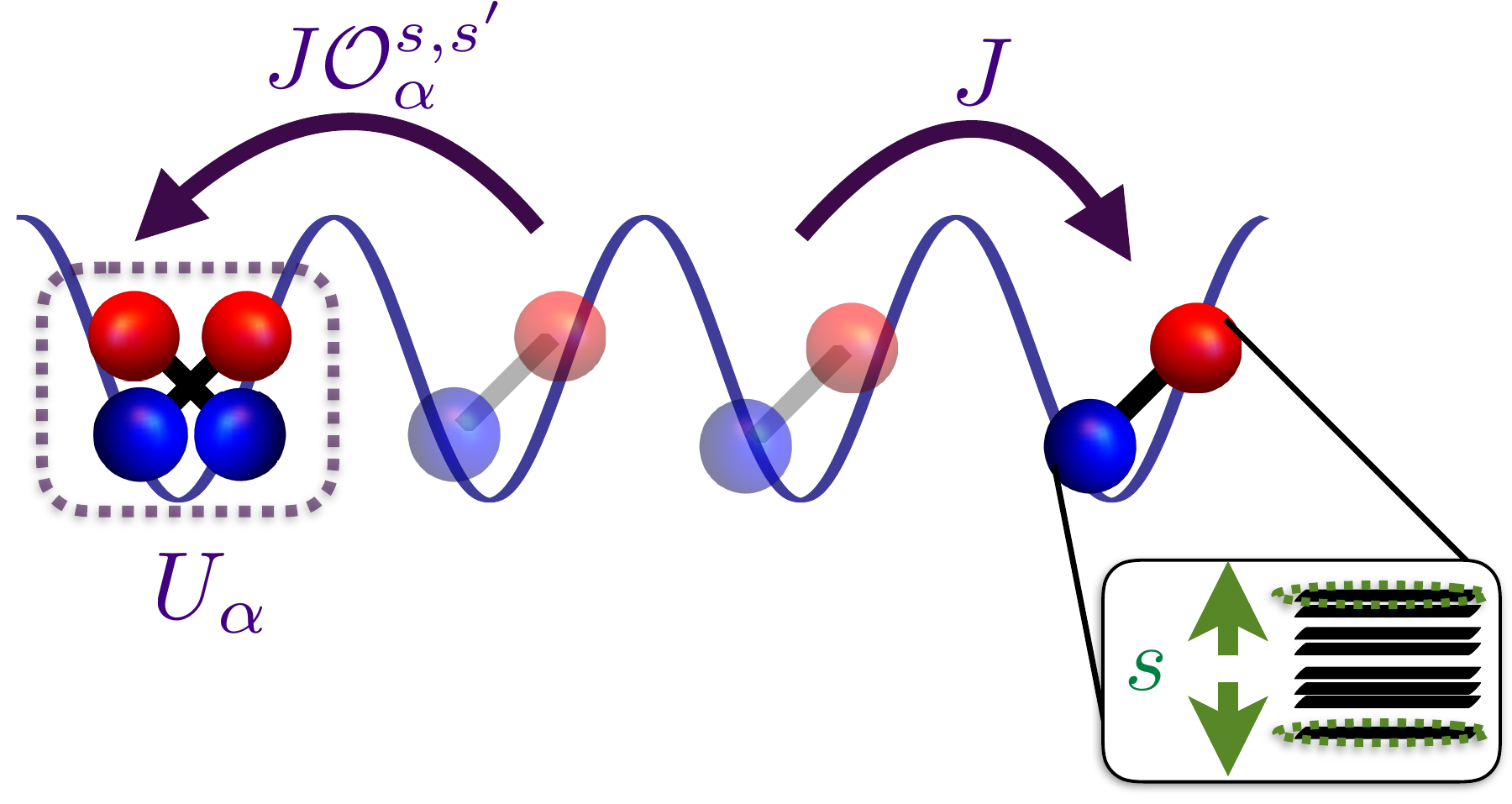}
\caption{ (Color online)   \emph{Schematic of effective model Eq.~\eqref{eq:EffectiveModel}}.  Ultracold NRMs in open-channel internal states $s$ tunnel through the lattice at a rate $J$ when moving to an empty site.  When an NRM tunnels onto an occupied site, the tunneling is reduced by the open-channel weight $\mathcal{O}_{\alpha}^{s,s'}$, the matrix element of two open channel molecules on the two-body on-site eigenstates $\ket{\alpha}$.  Energy shifts due to collisional resonances for two NRMs on a site are described by interaction energies $U_{\alpha}$.
\label{fig:latt-eff-model}
}
\end{figure}

Even though ultracold molecules have long been studied for their connections to quantum information~\cite{demilleD2002}, 
chemistry~\cite{doi:10.1021/cr300092g,0953-4075-39-19-S28,Balakrishnan2001652,PhysRevLett.115.063201,0034-4885-72-8-086401,PhysRevA.90.052716,krems2008cold}, and many-body physics~\cite{carr2009cold,doi:10.1021/cr2003568,wallrole,Gorshkov_Manmana_11,Gorshkov_Manmana_11b}, only recently has it 
been realized that nonreactive molecules (NRMs) in an optical lattice 
are described by an effective lattice model that is qualitatively modified from
 the conventional Hubbard model that governs their atomic counterparts~\cite{docaj:ultracold_2016}. The underlying physics is that when two 
molecules are close, there are many more configurations available than for two atoms: 
they can rotate and vibrate in many ways as they scatter off one 
other~\cite{mayle:statistical_2012,mayle:scattering_2013}.  These complex rotations and vibrations can alternatively be viewed in terms of bound eigenstates, the bimolecular collisional complexes (BCCs).  Ref.~\cite{docaj:ultracold_2016} derived the form of the 
``multi-channel Hubbard model" that governs ultracold NRMs in a lattice and estimated its parameters under a suite of approximations for the 
molecular scattering that were introduced by Refs.~\cite{mayle:statistical_2012,mayle:scattering_2013}.  This model introduces both a multi-channel on-site interaction and a channel-dependent tunneling for two molecules to reach the same site, as shown schematically in Fig.~\ref{fig:latt-eff-model}.

A major consequence of Ref.~\cite{docaj:ultracold_2016} 
is that all of the substantial literature that has studied NRMs in an optical lattice~\cite{1367-2630-11-5-055027,PhysRevA.82.013611,0034-4885-72-12-126401,PhysRevLett.104.125301,PhysRevLett.108.115301,0034-4885-78-6-066001,doi:10.1021/cr2003568} must be reconsidered in light of the modified on-site interaction: Rather than simply augmenting the normal on-site interaction (i.e.~``Hubbard $U$") with a dipolar interaction (for polar molecules), as has been done in the previous literature, a proper treatment must also include the multi-channel interaction.  
This on-site interaction term is not merely a small correction, but instead can potentially modify qualitative features of the physics such as the many-body phase diagram. It remains to be seen when and to what extent the modifications are significant, but we expect the effects to be substantial in many cases.

The present work is dedicated to a detailed derivation of the multi-channel lattice model of Ref.~\cite{docaj:ultracold_2016}, including generalizing this model to multiple internal states, e.g. hyperfine or rotational states.  Sec.~\ref{sec:overview} presents a broad view of the past and current work in ultracold molecules, including NRMs, with special focus on optical lattice settings.  This section also presents the many-body lattice model Eq.~\eqref{eq:EffectiveModel}, the microscopic derivation of which is our main result, along with a description of the terms appearing in it.  Further, this section briefly overviews few- and many-channel models in harmonic traps, which play a central role in our derivation.  Readers who are not interested in this background can skip to Sec.~\ref{sec:multi-channel-microscopic}.

Sec.~\ref{sec:multi-channel-microscopic} derives the effective resonance model, Eq.~\eqref{eq:H-2-body-ho} below, that governs two molecules on a single site of an optical lattice, where the two molecules couple to a dense collection of {BCC}s. This is a key result that is later used to obtain the lattice model, in Sec.~\ref{sec:latt-mod-derive}. In contrast to Ref.~\cite{docaj:ultracold_2016}, which posits the form of Eq.~(6), in Sec~\ref{sec:multi-channel-microscopic} we provide a formal derivation from a microscopic Hamiltonian that treats the two molecules as four pairwise interacting atoms.  Although this four-body problem is not straightforwardly solvable, even by advanced numerical methods, this section sets up a formalism that might allow these couplings to be computed with more advanced methods or future computational resources. It is also illuminating to see the relationship between the effective interaction parameters and conceptually simple microscopic expressions.

As is well known from the two-channel case, the energies in resonance models diverge as the coupling of open and closed channels approaches zero range.  Sec.~\ref{sec:regularization} derives in detail the regularization of the coupling constants that is necessitated to obtain finite, physical results when taking the couplings to zero-range. The form of this regularization has some unique features that are absent in the usual single- or two-channel cases, requiring the introduction of couplings between the bound states.  Then, Sec.~\ref{sec:latt-mod-derive} derives in detail the lattice model from the one-site two-particle solution.  We also derive extensions of the model to multiple internal  states  -- hyperfine, rotational, or vibrational.  Finally Sec.~\ref{sec:outlook} concludes and provides an outlook.

\section{Overview of nonreactive molecules in an optical lattice \label{sec:overview}}

Atomic physics has been transformed by the development of laser cooling, which uses closed cycling transitions to remove entropy from atoms via spontaneous emission.  Extending this technology directly to molecules is highly desirable, but early analyses~\cite{Stwalley} identified a limitation that persists to this day: due to the complex internal structure of molecular rotations and vibrations, excitation energy in a molecule can be distributed through many different pathways, rendering the existence of closed cycling transitions rare.  In atoms, the problem of hyperfine branching can be solved by adding repumping lasers. However, even for simple molecules, the number of lasers required makes this solution unsustainable aside from a few exceptional cases~\cite{di2004laser,shuman2010laser,1367-2630-17-3-035014,PhysRevLett.110.143001,1367-2630-17-5-055008,PhysRevLett.114.223003,PhysRevLett.116.063004}.

While direct cooling of molecules has proven to be challenging, many experiments have had success creating ultracold molecules via indirect methods.  Most prominently, there has been spectacular progress in ``assembling" ultracold molecules from a dual-species gas of pre-cooled alkali atoms. Here, the assembly occurs in two steps. First, one associates the atoms into a loosely bound Feshbach molecule by sweeping through a magnetically tunable Feshbach resonance. Next, one coherently transfers the molecular population to the rovibrational ground state using stimulated Raman adiabatic passage (STIRAP)~\cite{Shapiro}.   This final step is possible because of modern highly-coherent laser technology.  The first near-degenerate gas of molecules produced in the fashion was KRb~\cite{Ni_Ospelkaus_08}.

It was quickly learned that KRb, like half of the alkali dimers~\cite{PhysRevA.81.060703}, is chemically reactive through the pathway $\text{AB}+\text{AB} \to \text{A}_2+\text{B}_2$~\cite{Ospelkaus_Ni_10b}.  Since this initial demonstration, ultracold KRb molecules have enabled many fascinating studies~\cite{Ospelkaus_Peer_08,Ospelkaus_Ni_09,PhysRevLett.109.230403}, including phenomena such as cold collisions~\cite{deMiranda2011,Ospelkaus_Ni_10b}, suppression of chemical losses through trapping geometry~\cite{PhysRevLett.108.080405}, and the quantum Zeno effect~\cite{zhu_suppressing_2014}, and many-body physics for which chemical reactions are irrelevant, such as quantum spin models~\cite{yan_observation_2013,PhysRevLett.113.195302,wallrole}.  

Similar indirect molecule-production experiments have been successfully performed for the molecules RbCs~\cite{PhysRevA.85.032506,PhysRevA.89.033604,molony:creation_2014,takekoshi:ultracold-RbCs_2014,molony2016production,gregory2016controlling}, NaK~\cite{PhysRevLett.109.085301,park:two-photon_2015,park:ultracold_2015,park2016second}, and NaRb~\cite{1367-2630-17-3-035003,guo2016creation}. Unlike KRb, these are expected to be nonreactive.  In addition to the myriad experiments progressing with chemically reactive species~\cite{PhysRevA.86.021602,PhysRevA.89.020702,deiglmayr:formation-LiCs_2008,deiglmayr:permanent-LiCs_2010}, many experiments are underway attempting to cool molecules whose chemical reactivity is unknown~\cite{shuman2010laser,1367-2630-17-3-035014,PhysRevLett.110.143001,1367-2630-17-5-055008,PhysRevLett.114.223003,stuhl2012evaporative,zeppenfeld2012sisyphus,chervenkov2014continuous,PhysRevA.92.023404,lemeshko2013manipulation,KCsNagerl}.  Further details on the production of ultracold molecules through both direct and indirect means can be found in recent review articles~\cite{Meerakker, Hutzler, carr2009cold, KochandShapiro, Koehler,doi:10.1021/cr300092g,0953-4075-49-15-152002}.

The efficiency of the molecule formation process by way of magneto- or photo-association can be enhanced relative to free space by assembling the atoms into molecules in an optical lattice, where, ideally, exactly one atom of each species that is being combined would occupy a single lattice site~\cite{PhysRevLett.90.110401,PhysRevA.81.011605,PhysRevA.92.063416}.  Indeed, such an enhancement in the phase-space density of KRb molecules by assembling in an optical lattice has been recently observed~\cite{moses2015creation}.   Hence, optical lattices are a natural setting for high-density gases of ultracold molecules, including NRMs, in near-term experiments.  Not only are these systems cold, but the internal degrees of freedom can be controlled: once in the rovibrational ground state, further transfer to any desired hyperfine state, including the absolute ground state, can then be achieved with microwave control, owing to the mixing of hyperfine and rotational angular momenta by a nuclear quadrupole coupling~\cite{Ospelkaus_Ni_10,PhysRevLett.116.225306,gregory2016controlling}.  The use of an optical lattice is also arguably essential for many-body physics with reactive molecules; experiments must confine the molecules in particular geometries where chemical reactions are suppressed~\cite{PhysRevLett.105.073202,1367-2630-17-3-035007,1367-2630-17-1-013020} and work at times short compared to timescales of residual losses.  

In contrast to reactive molecules, NRMs do not suffer from geometrical and lifetime constraints, and so lead to exciting new possibilities for many-body physics in which translational motion of the molecules and the dipole-dipole interaction both are important.  Further, the possibility of having a large ratio of elastic to inelastic collisions leads to the possibility of evaporative cooling~\cite{zhu:evaporative_2013,stuhl2012evaporative} to bring molecules even deeper into the deep ultracold regime.  In order to harness these advantages of NRMs, however, we must understand the short-range collisional properties of NRMs and their implications for many-body lattice models.  

\begin{figure}[t]
\includegraphics[width=0.9\columnwidth]{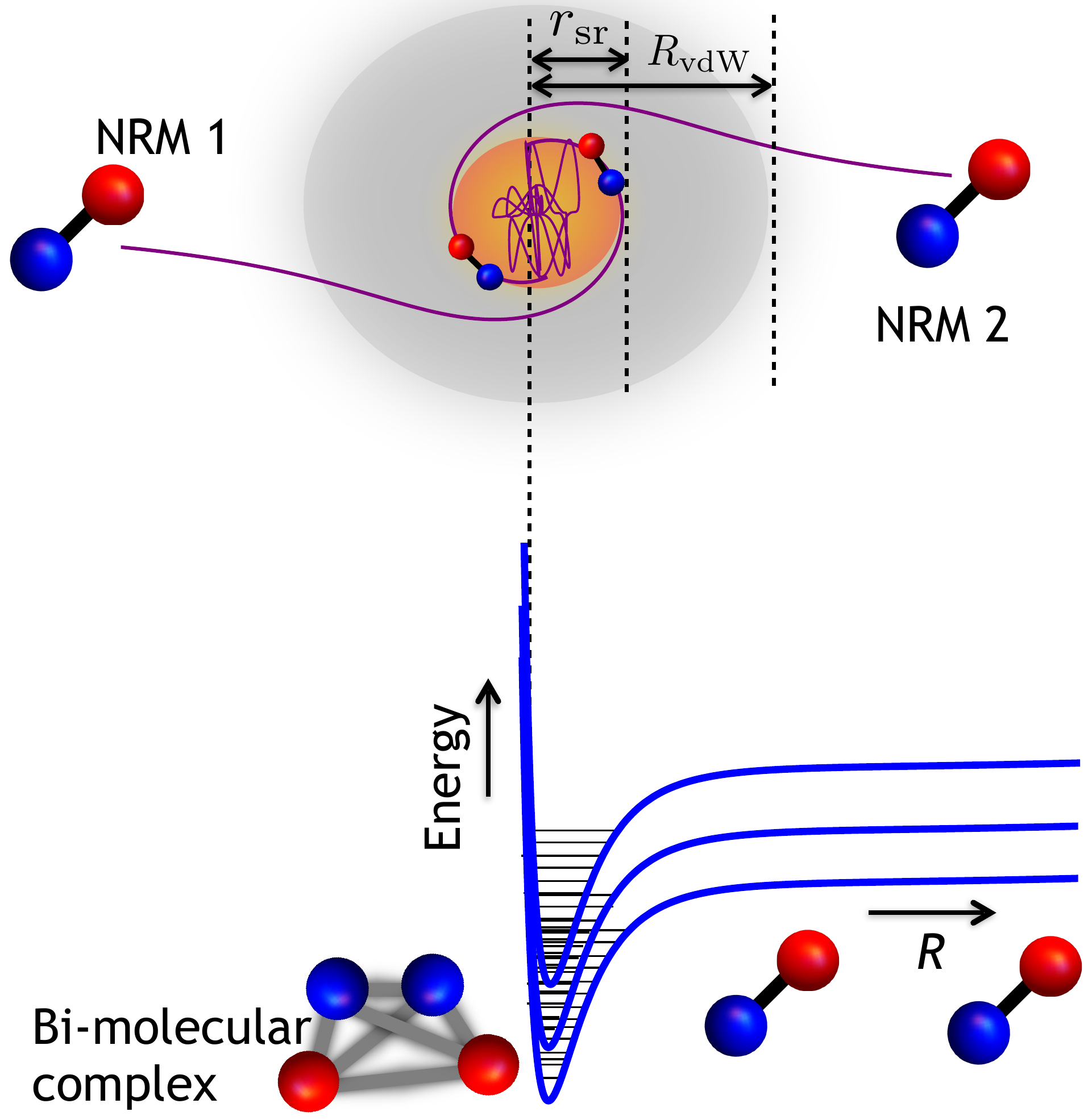}
\caption{ (Color online)   \emph{Complex Scattering of ultracold nonreactive molecules}.  At intermolecular separations large compared to a characteristic potential range $R_{\mathrm{vdW}}$, molecules propagate ballistically.  These ballistic trajectories are curved in a regular fashion in an intermediate range comparable to the potential length but still larger than a characteristic short-range length $r_{\mathrm{sr}}$.  Below $r_{\mathrm{sr}}$, many internal molecular states (e.g. rotations and vibrations) become strongly mixed as the constituent atoms undergo chaotic dynamics.  These complex dynamics can be recast in terms of a resonance model with a large density of rovibrational resonances, see Eq.~\eqref{eq:H-2-body-ho}.
\label{fig:molcmplx}
}
\end{figure}

Recently, it has been argued that ultracold collisions between NRMs are qualitatively different from those of alkali atoms due to a remarkably high density of internal states $\rho_b$, i.e.~rovibrational configurations, for the two molecule system occurring at small intermolecular separations~\cite{mayle:statistical_2012, mayle:scattering_2013,PhysRevA.89.012714}, see Fig.~\ref{fig:molcmplx}.  A typical value is 
\be 
\rho_b &\sim& \frac{1}{2\text{nK}} \sim  \frac{1}{2\pi \times 20\text{Hz}}\hspace{0.5in} \text{for NaK}
\ee
such that there are numerous scattering resonances within a typical thermal energy window even for the coldest $\gsim 100$nK NRMs.
(Note that $\hbar=k_B=1$ unless otherwise specified throughout.)
 These internal states lead to a near-continuum of resonances which will remain unresolved for realistic experimental temperatures, in contrast to atomic collisions, in which one or at most a few resonances are relevant at ultralow collision energies.  As a consequence, the standard approach for deriving effective lattice models for atoms based on a single-channel pseudopotential~\cite{jaksch_bruder_98,lewensteinM2007} will rarely apply to NRMs.

An effective model for NRMs based on this picture of a high density of resonances at zero energy was first presented in Ref.~\cite{docaj:ultracold_2016}.  In contrast to the Hubbard model which typically provides an accurate description for ultracold atoms in optical lattices, the effective model for NRMs takes the form of a multi-channel resonance model,
\begin{align}
\label{eq:EffectiveModel} \hat{H}=&-J\sum_{\langle i,j\rangle, s}\left[ \hat{c}^{\dagger}_{i,s}\hat{c}_{j,s}+\mathrm{H.c.}\right]+\sum_i\left(\sum_{\alpha} U_{\alpha} \hat{n}_{i,\alpha}+\frac{3\omega}{2}\hat{n}_i\right)\,,
\end{align}
whose derivation is outlined below and presented in detail in Sec.~\ref{sec:latt-mod-derive}.  This model, which generalizes the results of Ref.~\cite{docaj:ultracold_2016} to multiple open channel states, is valid for bosonic or fermionic NRMs in a deep lattice, subject to the constraint that no more than two NRMs occupy a single lattice site.  Here, $J$ is the tunneling, $U_\alpha$ is the interaction energy of a pair of NRMs in the state $|\alpha\rangle$, and $\omega$ is the harmonic trapping frequency of an NRM within an individual lattice site (see visualization in Fig.~\ref{fig:latt-eff-model}).  The states $|\alpha\rangle$ are the eigenstates of the relative coordinate Hamiltonian for two particles on one site [given by Eq.~\eqref{eq:H-2-body-ho} below].  The interaction energy $U_{\alpha}$ is found from the eigenenergy $E_{\alpha}$ of the state $|\alpha\rangle$ by  $U_{\alpha}= E_\alpha-3\omega/2$.  The index $s$ runs over the allowed states in the open channel, $\hat{n}_i$  counts the total number of NRMs on lattice site $i$, and $\hat{n}_{i,\alpha}$ measures the occupation probability of eigenstate $|\alpha\rangle$ on site $i$, respectively.  As an example, if NRMs can be in two states (say rotational or hyperfine) on different lattice sites, $s \in \{\uparrow,\downarrow\}$. This is controlled independently from the inevitable numerous rotational excitations of the BCC that contribute when two NRMs are on a single site, and which are indexed by $\alpha$.  In Ref.~\cite{docaj:ultracold_2016}, statistical distributions for the parameters appearing in this model, e.g., $U_{\alpha}$, were identified within a series of additional approximations.  A critical assay of these approximations, their possible breakdown, and more general theories will be presented in a separate work~\cite{wall:beyond_2016}.

The operators $\hat{c}_{i ,s}$ are bosonic or fermionic operators for bosonic or fermionic NRMs, i.e.~satisfy the usual (anti-)commutation relations between operators on different sites, but are modified from the usual annihilation operators to account for the many interaction channels and the low-filling constraint.  Explicitly, the actions of $\hat{c}^{\dagger}_{i,s}$ on the vacuum state $|0\rangle_i$, a lattice site with a single molecule in state $|s'\rangle_i$, and a site with two molecules in the relative state $|\alpha\rangle_i$ are
\begin{align}
\label{eq:c1}\hat{c}^{\dagger}_{i, s}|0\rangle_i&=|s\rangle_i\, ,\\
\hat{c}^{\dagger}_{i, s}|s'\rangle_i&=P_{s,s'}\sqrt{1+\delta_{s,s'}}\sum_{\alpha}\mathcal{O}^{s,s'}_{\alpha}|\alpha\rangle_i\, ,\\
\label{eq:c3}\hat{c}^{\dagger}_{i, s}|\alpha\rangle_i&=0\, .
\end{align}
The first line is the usual creation of a molecule on an empty site, and the last line is the low-filling constraint.  The center line is the creation of a superposition of two-body eigenstates $|\alpha\rangle_i$ by adding a molecule to an already occupied site.  The square-root term accounts for Bose stimulation when the two molecules are bosonic and in the same internal state, and the coefficients $P_{s,s'}$ account for fermionic exchange and Pauli blocking (or are all unity in the case of bosons).  See Sec.~\ref{sec:latt-mod-derive} for details.  Finally, the overlap $\mathcal{O}^{s,s'}_{\alpha}\equiv \langle \alpha|s,s'\rangle$ describes the projection of two open-channel molecules on a single site $i$ onto the set of two-body relative eigenstates $|\alpha\rangle_i$.  We stress that Eq.~\eqref{eq:EffectiveModel} makes no assumptions about the character of the states $|\alpha\rangle_i$, for example about the degree of lattice band-mixing, provided that all energy scales (e.g. temperature and interactions $U_{\alpha}$) are small compared to the band gap.

The $s$ in Eq.~\eqref{eq:EffectiveModel} that labels internal states in principle can refer to hyperfine, rotational, or vibrational states.  It can be readily generalized to other excitations, e.g., band excitations, by making $J$ state-dependent. Two conditions are required on internal states for this model to apply. First, they must be  long-lived on the timescales of the experiment (ruling out extremely highly excited rotational states, moderately excited vibrational states, and nearly all electronic states). Second, the energies of the internal states must be large compared to $J$ and $U_{\alpha}$: this ensures that the population of each internal component is separately preserved by forcing population changing interactions to be far off-resonant. These conditions are satisfied for hyperfine levels at the $\sim 100$G magnetic fields used for magnetoassociation of atoms into molecules.  

Most of these conditions are also easily satisfied for internal levels that correspond to low-lying rotational excitations of the molecules, and the form of Eq.~\eqref{eq:EffectiveModel} remains valid.  However, in this case, molecules in different rotational states on different sites can exchange rotational quanta through the dipole-dipole interaction, and such ``state-swapping" needs to be included in Eq.~\eqref{eq:EffectiveModel}.  Also, in the presence of an electric field, for polar molecules one must add long-ranged $1/R^3$ interactions sites separated by distance $R$~\cite{1367-2630-11-5-055027,PhysRevA.82.013611,0034-4885-72-12-126401,PhysRevLett.104.125301,PhysRevLett.108.115301,0034-4885-78-6-066001}.  The statistical probability distributions derived for the Hamiltonian parameters in Ref.~\cite{docaj:ultracold_2016} must be modified when higher rotational states are considered~\cite{wall:beyond_2016}.

The lattice Hamiltonian in Eq.~\eqref{eq:EffectiveModel} for the deep lattice is derived in two steps. First one computes the eigenstates for one  and two molecules in a single lattice site, approximated by an isotropic harmonic trap $V(r) = \frac{1}{2} m \omega^2 r^2 $, where $r$ is the displacement of the molecule from the trap center and $m$ the molecular mass.  The sites are then stitched together to determine the effective lattice  model. Sec.~\ref{sec:latt-mod-derive} describes this procedure.  For the single site system, the one-molecule solutions are the usual harmonic oscillator wavefunctions and energies. The two-molecule wavefunctions are solutions to the Hamiltonian  ${\hat H}={\hat H}_{\text{c.m.}}+{\hat H}_{\text{rel}}$, which separates in center-of-mass and relative coordinates for the harmonic trap. Here
${\hat H}_{\text{c.m.}}= \sum_{n_{\text{c.m.}},\ell_{\text{c.m.}}} (2n_{\text{c.m.}}+\ell_{\text{c.m.}}+3/2)\omega\ket{n_{\text{c.m.}},\ell_{\text{c.m.}}}\!\bra{n_{\text{c.m.}},\ell_{\text{c.m.}}}$, where ${ n}_{\text{c.m.}}$ and $\ell_{\text{c.m.}}$ are the principal and angular momentum numbers, $\ket{n_{\text{c.m.}},\ell_{\text{c.m.}}}$ is the corresponding center of mass eigenstate.  The ($s$-wave) relative coordinate Hamiltonian of two molecules in a harmonic trap can be written as
\be
\hat{H}_{\text{rel}} \!&=&\! \sum_n \epsilon_n \!\ket{n}\!\bra{n} + \sum_b \nu_b \! \ket{b}\!\bra{b} + \sum_{nb} \!\lp W_{nb}\! \ket{n}\!\bra{b}+\hc \rp \nonumber \\
\label{eq:H-2-body-ho}
\ee
with $\epsilon_n=(2n + 3/2) \omega$. The $\ket{b}$ are the short-range BCC bound states of the system and the $\ket{n}$ are harmonic oscillator states. 
 Under the assumption that  the oscillator length $\lho=\sqrt{1/(\mu\omega)}$, with $\mu=m/2$ the reduced mass (for two molecules, each of mass $m$), is large compared to the microscopic lengths that characterize the intermolecular interactions, the couplings simplify to 
\be 
W_{nb} &=& w_b M_n/\lho^{3/2} \label{eq:Wnb-factor}
\ee
for some set of $w_b$ that depend only on the BCC properties and with
\be
M_n &=& \sqrt{\frac{\Gamma(n+3/2)}{\Gamma(n+1)}} \label{eq:Mn-defn} 
\ee
where $\Gamma(x)$ is Euler's gamma function. An exact (though formal) derivation of this is discussed in Sec.~\ref{sec:multi-channel-microscopic}.  

One could calculate the $\nu_b$ and $w_{b}$, and from these calculate the $U_\alpha$ and ${\mc O}_{\alpha}^{s,s'}$ without approximation if one had highly accurate multi-channel interatomic interaction potentials between the atoms constituting the molecules. Even with such potentials, the resulting numerical calculation would be formidable. Although we won't solve the resulting equations, Sec.~\ref{sec:multi-channel-microscopic} formally sets up the equations necessary to carry this procedure out in principle. Furthermore, even without solving it, it allows us to identify that the \emph{form} of the effective model Eq.~\eqref{eq:EffectiveModel} is in principle exact and robust beyond the approximations used in Ref.~\cite{docaj:ultracold_2016}.  In addition to providing the foundation for future work, it serves  to clarify the meaning of the $\nu_b$s and $w_b$s.

\begin{figure}
 \includegraphics[width=1.0\columnwidth]{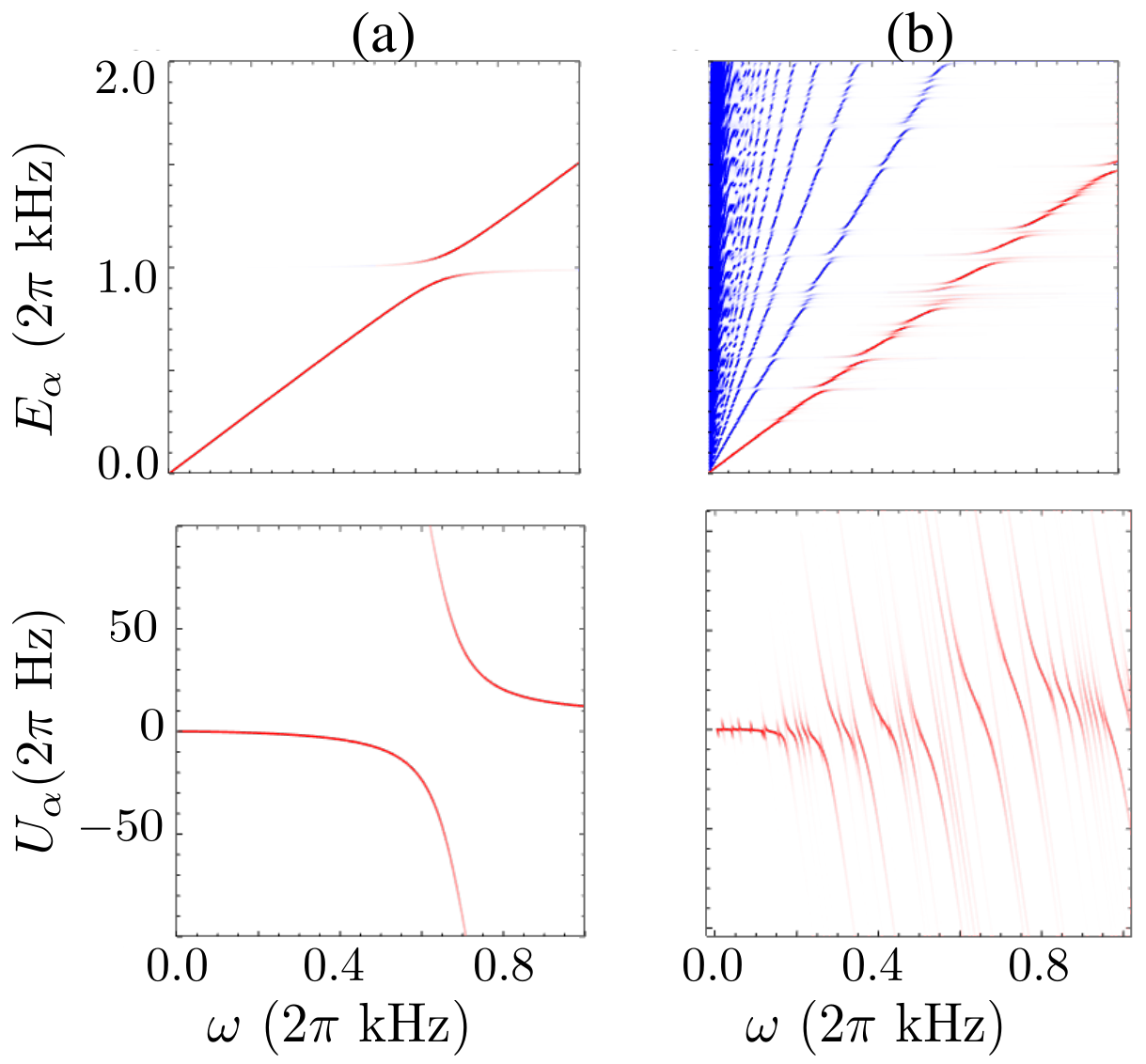}
\caption{ (Color online) \emph{Eigenvalues $E_\alpha$ (top row) and effective interactions $U_\alpha$ (bottom row) as a function of trap frequency $\omega$.} 
(a) Generic structure illustrated by a two-channel (single bound state) model. Top: The relative coordinate eigenenergy $E_\alpha$ shows an avoided crossing between a bimolecular collisional complex at fixed energy $2\pi \times 1.0$~kHz  and the harmonic oscillator ground state that is linearly dependent on $\omega$ (diagonal). 
The opacity at each plotted point (darkness of the line)  is set to be ${\mc O}_\alpha$ of the corresponding eigenstate,  indicating its importance to the lattice model. 
Bottom: On-site interaction in the lowest band (harmonic oscillator ground state), $U_\alpha$. The avoided crossing in $E_\alpha$ manifests as a strong, multivalued interaction near the crossing.
(b) $E_\alpha$ (top) and $U_\alpha$ (bottom) for realistic parameters for RbCs. Isolated resonances at small $\omega$ give way to coupled resonances at larger $\omega$. 
\label{fig:Oalpha-Ualpha}}
\end{figure}

Figure~\ref{fig:Oalpha-Ualpha} shows the eigenvalues $E_\alpha$ of Eq.~\eqref{eq:H-2-body-ho} and the lattice model parameters ${\mc O}_\alpha\equiv {\mc O}_{\alpha}^{1,1}$ and $U_\alpha$ as a function of $\omega$ for  a toy model with a single BCC [Fig.~\ref{fig:Oalpha-Ualpha}(a)] and for an NRM with realistic properties [Fig.~\ref{fig:Oalpha-Ualpha}(ab] (estimated for the bosonic NRM RbCs), as determined by the statistical framework of Ref.~\cite{docaj:ultracold_2016}. Although the precise form of the results depends on the details of the approximations of this statistical framework, the qualitative structure is expected to be robust against these approximations. 

Figure~\ref{fig:Oalpha-Ualpha}(a) shows the building block that is key to understanding all of the relevant structure seen in the model parameters. It shows results for a two-channel model, i.e.~a model with a single bound state, and focuses on the ground harmonic oscillator state. Because the bound state is so tightly confined, its relative-coordinate energy is independent of the trap and hence $\omega$. Absent coupling to the open channels (i.e., for $W_{nb}=0$), it would form a horizontal line, whose value in Fig.~\ref{fig:Oalpha-Ualpha}(a) is chosen to be $2\pi \times 1.0\text{kHz}$ for illustration. The harmonic oscillator relative-coordinate ground state energy  $3\omega/2$, gives the diagonal line. The coupling $W_{0b}$ between these two states leads to the avoided crossing structure in $E_\alpha$ that is observed in Fig.~\ref{fig:Oalpha-Ualpha}(a)~[top]. The contribution of the eigenstate to the lattice model is given by its overlap ${\mc O}_\alpha$ with the harmonic oscillator state.  The relevant eigenstate's energy is predominantly on the diagonal line except close to the resonance. Close to the resonance both states become important until they are equally weighted at resonance. 
The result for $U_\alpha$  follows from this: The $U_\alpha$ is the deviation of the eigenenergy $E_\alpha$ from the non-interacting energy (harmonic oscillator ground state). The $U_\alpha$ is close to zero for the single relevant state far from the resonance. Near the resonance, two $U_\alpha$ are important, with values $\pm W_{0b}/2$.

Turning to the more general case, Fig.~\ref{fig:Oalpha-Ualpha}(b) shows results for parameters estimated for a typical NRM RbCs. At small $\omega$, a series of isolated resonances exists, with the behavior near each similar to the two-channel cases. At larger $\omega$ these resonances smear together.

\section{Microscopic calculation of $w_b$ and $\nu_b$ from interatomic potentials\label{sec:multi-channel-microscopic}}

In principle, the couplings $w_b$ and energies $\nu_b$ can be calculated from a solution to the four-atom Schr\"odinger equation.   While such a calculation remains a technical challenge which we do not pursue here, the derivation itself helps to develop an intuitive understanding for the microscopic origin of these parameters, and may some day lead to a real solution of the problem.  Our goal here is two-fold:  first, to derive Eq.~\eqref{eq:H-2-body-ho} from a coordinate space representation of the four-atom system and give expressions for $w_b$ and $\nu_b$ in terms of four-atom wave functions in the adiabatic hyperspherical representation~\cite{MacekJPB1968}; second, to justify the factorization of the coupling $W_{nb}=w_{b}M_n/\lho^{3/2} $. For simplicity, the current treatment considers only the spatial degrees of freedom associated with the atoms and quantum numbers associated with the electronic or spin degrees of freedom are suppressed. It can, however, be readily generalized to include atomic hyperfine structure, molecules with more than two atoms, or to an even lower-level description in terms of nuclei and electrons. Readers familiar with methods in few-body physics will recognize equations~\eqref{jacobicoors}-\eqref{Upoteff} as standard material~\cite{MacekJPB1968,Lin1995PhysRep,rittenhouse2011hyperspherical} that is included here to provide the necessary background for what follows.

\subsection{Derivation of the two-body Hamiltonian Eq.~\ref{eq:H-2-body-ho}}
Our purpose here is to demonstrate how the multichannel molecule-molecule Hamiltonian Eq.~\eqref{eq:H-2-body-ho} emerges from a four-atom ($N=4$) problem in the limit where the intermolecular separation is large compared to the dimer size.  In doing so, the molecule-molecule interaction channels become unambiguously specified, and a formula for the channel couplings $W_{nb}$ is obtained. The reduction to a multichannel model is facilitated by expressing the positions $\{\boldsymbol{r}_i\}$ of the atoms in a set of mass-scaled ``H-type'' Jacobi coordinates $\{\boldsymbol{\rho}_i\}$:
\begin{eqnarray}
\label{jacobicoors}
\nonumber&\boldsymbol{\rho }_1=\sqrt{\frac{\mu_{1,2}}{\mu_{\text{4B}}}}\left(\boldsymbol{r}_1-\boldsymbol{r}_2\right)\, , \\
\nonumber&\boldsymbol{\rho }_2=\sqrt{\frac{\mu_{3,4}}{\mu_{\text{4B}}}}\left(\boldsymbol{r}_3-\boldsymbol{r}_4\right)\, , \\
&\boldsymbol{\rho}_3=\sqrt{\frac{\mu_{12,34}}{\mu_{\text{4B}}}}\left(\frac{m_3\boldsymbol{r}_3+m_4\boldsymbol{r}_4}{m_3+m_4}-\frac{m_1\boldsymbol{r}_1+m_2\boldsymbol{r}_2}{m_1+m_2}\right)\, , \\
\nonumber&\boldsymbol{X}=\frac{m_1\boldsymbol{r}_1+m_2\boldsymbol{r}_2+m_3\boldsymbol{r}_3+m_4\boldsymbol{r}_4}{m_1+m_2+m_3+m_4}\, .
\end{eqnarray}
Here, $\mu_{i,j}=m_im_j/(m_i + m_j)$ is the reduced mass of particles (or clusters) $i$ and $j$.  The mass scale $\mu_{\text{4B}}$ is arbitrary, but is often chosen as $\mu_{\text{4B}} = \left(\mu_{1,2}\mu_{3,4}\mu_{12,34}\right)^{1/3}$ in order to preserve the integration volume element~\cite{Delves:1960p6469}.  Note that $\mu_{12,34}$ coincides with $\mu=m/2$ introduced in~\cite{docaj:ultracold_2016} and in Section~\ref{sec:overview} above.  In a harmonic trapping potential, the center of mass motion decouples completely from the relative motion, allowing one to write $\Psi = \Psi_{\text{rel}}(\boldsymbol{\rho}_1,\boldsymbol{\rho}_2,\boldsymbol{\rho}_3)\Psi_{\text{c.m.}}(\boldsymbol{X})$.  
The remaining $d=3N-3=9$ relative coordinates may be transformed to hyperspherical coordinates: $\{\boldsymbol{\rho}_1,\boldsymbol{\rho}_2,\boldsymbol{\rho}_3\}\rightarrow \{R,\Omega\}$, where $R$ is the hyperradius, defined as
\begin{equation}
R^2 = \rho_1^2 + \rho_2^2 + \rho_3^2\, ,
\end{equation}
and $\Omega$ collectively denotes all of the eight remaining relative angular coordinates. This transformation can be made by defining canonical hyperangles, as reviewed for example in~\cite{SmirnovShitikova}, or through the introduction of so-called democratic hyperangles~\cite{PackParker1987JCP,aquilanti1997qmh, Kuppermann1997ea, rittenhouse2011hyperspherical}, which are capable of treating all fragmentation channels on an equal footing. Because the hyperradius $R$ is invariant with respect to particle permutations, it proves to be a more convenient collision coordinate than the intermolecular separation.  A significant conceptual advantage afforded by this choice is that all exchange symmetry can be incorporated into the hyperangular channel functions by appropriate boundary conditions on the hypersphere, with no need for additional atom-exchange interactions in the four-atom potential energy surface.   Note that if the size of each dimer is negligible in comparison to the intermolecular separation, then the hyperradius is approximately equal to the molecular separation.

The bimolecular Hamiltonian for the relative degrees of freedom in hyperspherical coordinates can then be expressed as
\begin{align}
\label{Hrel}
\nonumber \hat{H}_{\text{rel}}=&\frac{-1}{2\mu_{\text{4B}}R^{d-1}}\diff{}{R}\left(R^{d-1}\diff{}{R}\right) + \frac{1}{2}\mu_{\text{4B}}\omega^2 R^2 \\
& + \frac{\hat{\Lambda}^2(\Omega)}{2\mu_{\text{4B}}R^2} + \hat{V}(R,\Omega).
\end{align}
Here, $\hat{\Lambda}^2/(2\mu_{\text{4B}}R^2)$ represents the hyperangular (fixed $R$) kinetic energy, where $\hat{\Lambda}$ is the hyperangular momentum~\cite{AveryHypersphericalHarmonics}.  The harmonic oscillator potential is purely hyperradial, and $\hat{V}(R,\Omega)$ contains all interatomic interactions.  The particular form of $\hat{V}(R,\Omega)$ is not important for our purpose, except that it depends only on relative degrees of freedom.  We represent eigenstates $\Psi_{\text{rel}}(R,\Omega)$ of $\hat{H}_{\text{rel}}$ by writing
\begin{equation}
\label{psirel}
\Psi_{\text{rel}}(R,\Omega) = \sum_{\alpha}{F_\alpha(R)\Phi_\alpha(R;\Omega)},
\end{equation}
where the channel functions $\Phi_\alpha$ are defined as eigenstates of the adiabatic Hamiltonian
\begin{equation}
\hat{H}_{\text{ad}} = \frac{\Lambda^2}{2\mu_{\text{4B}}R^2} + \hat{V}(R,\Omega),
\end{equation}
with $R$-dependent adiabatic potentials $\mc{U}_\alpha(R)$ as eigenvalues:
\begin{equation}
\label{adSE}
\hat{H}_{\text{ad}}(R;\Omega)\Phi_\alpha(R;\Omega) = \mc{U}_\alpha(R)\Phi_\alpha(R;\Omega).
\end{equation}

Inserting Eq.~(\ref{psirel}) into the eigenequation for Eq.~\eqref{Hrel} and making use of Eq.~(\ref{adSE}) leads to a set of coupled channel equations in $R$ for $F_\alpha$,
\begin{widetext}
\begin{equation}
\label{HRCE}
\sum_{\beta} {\left[ \left(\frac{-1}{2\mu_{\text{4B}}}\secder{}{R} + \frac{1}{2}{\mu_{\text{4B}}\omega^2 R^2}+\mc{U}_\alpha^{\text{eff}}(R)-E \right) \delta_{\alpha\beta} - \frac{1}{2\mu_{\text{4B}}} \left(2 P_{\alpha\beta}(R)\der{}{R}+(1-\delta_{\alpha\beta})Q_{\alpha\beta}(R)\right) \right]  R^{(d-1)/2} F_\beta(R) }=0\, .
\end{equation}
\end{widetext}
The factor of $R^{(d-1)/2}$ is present to facilitate the removal of first-derivative terms in the radial kinetic energy. The effective potential in each channel becomes:
\begin{equation}
\label{Upoteff}
\mc{U}_{\alpha }^{\text{eff}}(R)=\mc{U}_{\alpha }(R) - \frac{{Q}_{\alpha \alpha }(R)}{2 \mu_{\text{4B}}} + \frac{(d-1)(d-3)/4}{2\mu_{\text{4B}} R^2}.
\end{equation}
It is critical to include the positive diagonal contribution $-{Q}_{\alpha \alpha}(R)/(2\mu_{\text{4B}})$ in order to obtain the correct large $R$ behavior of $\mc{U}_\alpha^{\text{eff}}$, which must approach the binding energy of two separated NRMs in their rovibrational ground state.   
The channels are coupled by non-adiabatic first-derivative and second-derivative couplings, defined as
\begin{equation}
\label{PmatQmat}
P_{\alpha  \beta}(R)=\left\langle \Phi _{\alpha}\left|\frac{\partial \Phi _{\beta }}{\partial R}\right.\right\rangle\, , \;\;\;\;\; 
Q_{\alpha\beta}(R) = \left\langle \Phi_\alpha\left|\secdiff{\Phi_\beta}{R}\right.\right\rangle\, ,
 \end{equation}
where the integration in these matrix elements is carried out over angular coordinates only.  Because the channel functions are orthonormal at each $R$, $\diff{}{R}\langle\Phi_\alpha|\Phi_\beta\rangle = 0$ immediately gives that $P_{\alpha\beta}$ is antisymmetric, and thus zero along the diagonal.  The functions $F_\alpha(R)$ can be expanded in a complete set of states $\psi_\alpha^i$ for each $\alpha$ as
\begin{equation}
\label{dDto3D}
R^{(d-1)/2}F_\alpha(R)=\sum_i{R \psi_\alpha^i(R)}\, .
\end{equation}
It is then straightforward to show that the Hamiltonian operator $\hat{H}_{\text{rel}}$ can be written in the form
\begin{equation}
\hat{H}_{\text{rel}}=\sum _{\alpha ,i} \sum _{\beta ,j} {|\psi _{\alpha }^i \alpha \rangle  \left(\delta _{\alpha \beta } \mathcal{H}_{\beta \beta }^{ij}+\mathcal{X}_{\alpha \beta }^{ij}\right) \langle \psi _{\beta }^j \beta |}\, ,
\end{equation}
where we define the diagonal matrix element $\mc{H}_{\beta \beta}^{ij}$:
\begin{align}
\nonumber &\mathcal{H}_{\beta \beta }^{ij}=\int_{0}^{R_m} \Big[\frac{1}{2 \mu_{\text{4B}}}\frac{d\psi _{\beta }^i}{dR} \frac{d\psi _{\beta }^j}{dR} \\ 
& + \psi _{\beta }^i\left(\mc{U}_{\beta }^{\text{eff}}(R)+\frac{1}{2}\mu_{\text{4B}}\omega^2 R^2\right) \psi _{\beta }^j \Big] \, R^2dR\, ,
\end{align}
and the channel coupling matrix elements as
\begin{align}
\label{Xcoupling}
\nonumber &\mathcal{X}_{\alpha  \beta}^{ij} = -\frac{1}{2 \mu_{\text{4B}}} \int_{0}^{R_m} \Big[P_{\alpha \beta }(R) \Big(\psi _{\alpha }^i \frac{d\psi _{\beta }^j}{dR}-\psi _{\beta }^j \frac{d\psi _{\alpha }^i}{dR}\Big)\\
&-\left(1-\delta _{\alpha \beta }\right) \psi _{\alpha }^i \psi _{\beta }^j {\tilde{Q}}_{\alpha \beta }(R)\Big] \, R^2dR\, .
\end{align}
In order to demonstrate that $\hat{H}_{\text{rel}}$ is explicitly symmetric, we have introduced the symmetric form of the second derivative coupling
\begin{equation}
\label{eq:Qtil}
{\tilde{Q}}_{\alpha \beta }(R)=\left\langle\diff{\Phi_\alpha}{R} \right.\left|\diff{\Phi_\beta}{R}\right\rangle\, ,
\end{equation}
which is related to the matrices ${P}$ and ${Q}$ by  $\tilde{{Q}} = \diff{{P}}{R} - {Q}$.

Ultimately, to derive Eq.~\eqref{eq:H-2-body-ho}, a rigorous connection between the $d=9$ dimensional four-atom space and the effectively 3D molecule-molecule channel is needed.  To enable such a connection, we first focus on the solutions to Eq.~\eqref{HRCE} ignoring all off-diagonal elements $\alpha\ne\beta$, which we denote $f_\alpha(R)$. First, we write $\mc{U}_\alpha(R)-Q_{\alpha\alpha}(R)/2\mu_{\text{4B}} = \lambda(\lambda+d-2)/2\mu_{\text{4B}} R^2 $, where $\lambda$ is assumed to be in general a function of $R$, and independent of $R$ only for the noninteracting case where it is equal to the hyperangular momentum eigenvalue of the $\hat{\Lambda}^2$ operator~\cite{AveryHypersphericalHarmonics}.  It is convenient to define an ``effective'' hyperangular momentum $\ell_{\text{eff}} = \lambda + (d-3)/2$ such that the uncoupled radial function $f_\alpha(R)$ satisfies
\begin{align}
\label{eq:radialHO}
\Big[&\frac{-1}{2\mu_{\text{4B}}}  \secder{}{R} +\frac{1}{2}\mu_{\text{4B}}\omega^2 R^2\\
\nonumber &  + \frac{ (\ell_{\text{eff}}+1/2)^2-1/4 }{2\mu_{\text{4B}} R^2} - E \Big] R^{(d-1)/2}f_\alpha(R)=0\, .
\end{align}
The solution for {\em constant} $\ell_{\text{eff}}$ is~\cite{AbramowitzStegun}
\begin{equation}
\label{dDimho}
\tilde{R}^{\frac{(d-1)}{2}}f_{n,\ell_{\text{eff}}}(\tilde{R}) = \sqrt{ \frac{2n!}{\Gamma(n+\frac{3}{2})}} \tilde{R}^{\ell_{\text{eff}}+1}e^{-\frac{\tilde{R}^2}{2}}  L_n^{\ell_{\text{eff}}+\frac{1}{2}}(\tilde{R}^2)\, ,
\end{equation}
where $\tilde{R}=R/\sqrt{1/\mu_{\text{4B}}\omega}$.  For $s$-wave scattering in any of the two-body channels, we expect that as $R\gg\rvdw$, $[(\ell_{\text{eff}}+1/2)^2-1/4]/2\mu_{\text{4B}} R^2 \rightarrow E_{\text{mm}}$, where $E_{\text{mm}}$ is the threshold energy of the molecule-molecule channel.  Thus, Eq.~\eqref{eq:radialHO} for constant $\ell_{\text{eff}}$ is reduced to an $s$-wave harmonic oscillator equation whose solutions, Eq.~\eqref{dDimho}, have eigenvalues $E_n=E_{\text{mm}}+\omega(2n+3/2)$, and are simply related to the 3D $s$-wave oscillator eigenfunctions, $\psi_{\ell}^n$, by $R^{(d-1)/2}f_{n,\ell_{\text{eff}}=0}(R)=R \psi_{\ell=0}^n(R)$.  These oscillator functions are precisely the basis functions chosen to represent the open channel wavefunction component.  They are eigenstates of the open channel problem in the absence of coupling to bound states, and in the absence of an open-channel resonance.  It is convenient to measure their eigenenergy with respect to the molecule-molecule threshold by writing
\begin{equation}
\label{Eho}
\epsilon_n=E_n-E_{\text{mm}}=\omega(2n+3/2).
\end{equation}  

To proceed, let $O$ denote the open channel, and $B$ denote a closed channel.  The open-channel $s$-wave oscillator basis states described above are now denoted $|\psi _O^n\rangle$. Basis states in each closed channel are chosen to be the set of bound eigenstates $|\psi_B^a\rangle$ of the single-channel Hamiltonian $\hat{H}_B= \sum_{i,j}|\psi _B^i B\rangle \mathcal{H}_{BB}^{ij} \langle \psi _B^j B|$ with eigenvalue $\lambda_B^a$.  That is, we let
\begin{equation}
|\psi _{\alpha }^j \alpha \rangle =
\begin{cases}
 |\psi _B^a B\rangle  & \alpha \in B \\
 |\psi _O^n O\rangle  & \alpha \in O \\
\end{cases}\, ,
\end{equation}
where $\hat{H}_B|\psi_B^a\rangle = \lambda_B^a|\psi_B^a\rangle$.  The states $|\psi_O^n\rangle$ would have difficulty capturing the short-range boundary condition in the presence of a shallow bound state in the open channel. In that case, one could alternatively use the known solutions for a contact interaction in a harmonic trap~\cite{Busch_Englert_98}.  

With only one open channel, the Hamiltonian can be written explicitly in terms of open and bound channels as
\begin{align}
\label{Hrel2}
\nonumber \hat{H}_{\text{rel}}=&\sum_n |\psi _O^nO\rangle \epsilon _n\langle \psi _O^nO|+\sum _{B a} |\psi_B^aB\rangle \lambda _B^a\langle \psi_B^aB| \\
\nonumber&+\sum _n \sum_{B, a} \left(|\psi _O^nO\rangle \mathcal{X}_{O B}^{n a}\langle \psi_B^aB| + \hc\right)\\
&+\sum_{B,B',a,a'} |\psi_B^aB\rangle \mathcal{X}_{B B'}^{a a}\langle \psi_{B'}^{a'}B'|\, .
\end{align}
Equation~\eqref{Hrel2} defines the various matrix elements in terms of open and closed channel basis states.
The states $|b\rangle$ appearing in Eq.~\eqref{eq:H-2-body-ho} are obtained by diagonalizing the subspace of bound-bound channels, $\hat{H}_{\text{b}}|b\rangle = \nu_b |b\rangle$, where
\begin{equation}
\label{eq:Hb}
\hat{H}_{\text{b}}=\sum_{a B}{|\psi_B^a B\rangle \lambda _B^a \langle \psi_B^a B|} + \sum_{B,B',a,a'} {|\psi_B^a B\rangle  \mathcal{X}_{B B'}^{a a'} \langle \psi _{B'}^{a'} B'|} \, .
\end{equation}
One can write the resulting eigenstates as $|b\rangle =\sum_{a B}{\langle\psi_B^aB|b\rangle |\psi_B^a B\rangle }$.
Rotating into the basis of eigenstates of $\hat{H}_{\text{b}}$, the block structure of the Hamiltonian becomes
\begin{align}
\nonumber \hat{H}_{\text{rel}}=&
\left(
\begin{BMAT}(b){c:ccc}{c:ccc} 
O O & O B_1 & O B_2 & \cdots \\
B_1 O & B_1 B_1 & B_1 B_2 & \cdots \\
B_2 O & B_2 B_1& B_2 B_2  & \cdots \\
\vdots & \vdots & \vdots  & \ddots
\end{BMAT}
\right) 
\\
&\to\left(
\begin{BMAT}(b){c:ccc}{c:ccc}
\epsilon_n &  \cdots & W_{nb} & \cdots \\
\vdots & \ddots & 0 & 0 \\
W_{bn} & 0 & \nu_b & 0 \\
\vdots & 0 & 0 & \ddots
\end{BMAT}
\right)\, .
\end{align}
The energies $\nu_b$ and the couplings $W_{nb}$ are now given explicitly in terms of the eigenstates of the microscopic Schr\"odinger equation as
\begin{align}
\label{eq:nubexact}
&\nu_b =\sum_{B,B^\prime,a,a^\prime}{\langle b|\psi_B^aB\rangle\langle\psi_{B^\prime}^{a^\prime}B^\prime|b\rangle \left(\lambda_B^a \delta _{a B,a^\prime B^\prime}+\mathcal{X}_{B B^\prime}^{a a^\prime}\right)}\, ,\\
\label{eq:Wnb1}
&W_{n b}=\sum_{B,a}{\langle \psi_B^aB|b\rangle \mathcal{X}_{O B}^{n a}}\, ,
\end{align}
such that with the short-hand $|n\rangle = |\psi_O^n O\rangle$, we recover Eq.~\eqref{eq:H-2-body-ho}
\begin{equation}
\hat{H}_{\text{rel}}=\sum_{n,b} \left(|n\rangle  W_{n b} \langle b| +\hc \right)+\sum_b |b\rangle \nu_b \langle b| +\sum_n |n\rangle  \epsilon_n\langle n| \, .
\end{equation}

\subsection{Factorization of $W_{nb}$}
\label{sec:Wnbfactor}
Finally, we sketch how the separation of length scales $l_b \ll l_{\text{ho}}$, with $l_b$ the characteristic length scale of the bound state, allows for the factorization of $W_{nb}=w_b M_n/l_{\text{ho}}^{3/2}$.  We focus on the channel coupling elements between the open and closed channels,
\begin{align}
\label{XOB}
\nonumber \mc{X}_{O B}^{na}=&\frac{-1}{2\mu_{\text{4B}}}\int_0^{R_m}\Big[ P_{OB}\left( \psi_O^n\frac{d\psi_B^a}{dR}-\psi_B^a \frac{d\psi_O^n}{dR} \right)\\
& - \tilde{Q}_{OB}\psi_O^n\psi_B^a\Big]R^2dR\, .
\end{align}
The closed channel eigenfunctions $\psi_B^a$ are all short ranged with support only over length scales $R\sim l_b$, and exponentially small for $R\gsim l_b$.  On the other hand, the oscillator state $\psi_O^n(R)$ is a long-range function that varies over length $l_{\text{ho}}/\sqrt{n}$.  

In order to exploit this separation of length scales, we first note that at short range, we may approximate the slowly varying oscillator state as $\psi_O^n(R)\approx \psi_O^n(0) + \left[\frac{d\psi_O^n}{dR}\right]_0 R$. Using Eq.~\eqref{dDimho} with $\ell_{\text{eff}}=0$, along with the fact that we have chosen our open-channel basis functions to be $R\psi_O^{n} = R^{(d-1)/2}f_{n,\ell_{\text{eff}}=0} $, we find
\begin{equation}
\label{psiho0}
\psi_O^n(0)=\sqrt{\frac{8\;\Gamma(n+3/2)}{\lho^3 \pi\;\Gamma(n+1)}} \;\;\;\;\;\; \left[\diff{\psi_O^n}{R}\right]_{R=0}=0
\end{equation}
Because $\psi_B^a(R)$ has support only for $R\sim l_b$, the term in Eq.~\eqref{XOB} containing $\diff{\psi_O^n}{R}$ is negligible compared to the other terms, and Eq.~\eqref{XOB} is cast into the form
\begin{equation}
\label{XOB2}
\mc{X}_{O B}^{na}=\frac{-1}{2\mu_{\text{4B}}}\psi_O^n(0) \int_0^{R_m} { \mc{I}_{OB}^a(R) R^2 dR} \, ,
\end{equation}
where we have defined
\begin{equation}
\mc{I}_{OB}^a(R)= P_{OB}(R)\left(\frac{d}{dR}-\tilde{Q}_{OB}(R)\right)\psi_B^a(R)
\end{equation}
Alternatively, one could perform an integration by parts on the second term in Eq.~\eqref{XOB}, removing the derivative on $\psi_O^n$.  The resulting surface term vanishes, and using the relation $\tilde{{Q}} = \diff{{P}}{R} - {Q}$, one obtains an equivalent expression for $\mc{I}_{OB}^a$,
\begin{equation}
\mc{I}_{OB}^a(R)= \left[ 2 P_{OB}(R)\left( \frac{d }{dR}+ \frac{1}{R}\right) +  Q_{OB}(R)\right]\psi_B^a(R) \, .
\end{equation}
The function $\mc{I}_{OB}^a(R)$ has support only at length scales $R\sim l_b$. Finally, the couplings $W_{nb}$ given in Eq.~\eqref{eq:Wnb1} are written as
\begin{equation}
\label{eq:obwbexact}W_{n b}\approx \psi_O^n(0) \sum_{B,a}{ \langle \psi_B^aB|b\rangle \frac{-1}{2\mu_{\text{4B}}}
  \int_0^{R_m}{\mc{I}_{OB}^a(R) R^2dR}}\, .
\end{equation}
Using Eq.~\eqref{psiho0}, one recovers Eqs.~\eqref{eq:Wnb-factor}-\eqref{eq:Mn-defn}, such that the couplings $w_b$ are now given as
\begin{equation}
\label{eq:wbexact}w_b = \frac{-\sqrt{2}}{\sqrt{\pi}\mu_{\text{4B}}} \sum_{B,a} { \langle \psi_B^aB|b\rangle \int_0^{R_m}{\mc{I}_{OB}^a(R)}R^2\,dR}\, .
\end{equation}
Were we to include an open channel resonance by replacing the oscillator states $\psi_O^n$ with the known two-body solutions for zero-range interactions~\cite{Busch_Englert_98}, then the $\epsilon_n$s and $M_n$s would be shifted, however the $w_b$s, which are independent of open-channel states, would be unchanged.

To summarize, in order to determine the energies $\nu_b$, one would first calculate the adiabatic channel functions $\Phi_\alpha(R;\Omega)$ and potentials $\mc{U}_\alpha(R)$ by solving the eigenvalue problem Eq.~\eqref{adSE}, then calculate the non-adiabatic couplings $P_{\alpha\beta}(R)$ and $\tilde{Q}_{\alpha\beta}(R)$ by Eqs.~\eqref{PmatQmat} and \eqref{eq:Qtil}.  One could then calculate the couplings $\mc{X}_{\alpha\beta}^{ij}$ by performing the integrals in Eq.~\eqref{Xcoupling} before finally diagonalizing $\hat{H}_{\text{b}}$ given in Eq.~\eqref{eq:Hb} to find its eigenvalues, $\nu_b$.  Diagonalizing Eq.~\eqref{eq:Hb} also yields the overlaps $\langle \psi_B^aB|b\rangle$ that appear both in Eq.~\eqref{eq:Wnb1} for the couplings $W_{nb}$, and in Eq.~\eqref{eq:wbexact} for the factorized coupling $w_b$.  In this way, all of the parameters in the zero-range model Eq.~\eqref{eq:H-2-body-ho} may be in principle determined from a microscopic theory.
 
Finally, we note that the adiabatic hyperspherical representation employed above, while conceptually illuminating, suffers from some shortcomings that may make a practical calculation difficult.  In practice, the hyperradial derivatives in Eq.~\eqref{PmatQmat} are usually estimated by a three-point difference rule.  In cases where the potential curves exhibit sharp avoided crossings, the couplings are sharply peaked, and difficult to calculate accurately.  The method of ``slow variable discretization''~\cite{TolstikhinJPB1996} (SVD) circumvents the calculation of hyperradial derivatives, and has been shown to provide fast convergence for three-body systems with many two-body bound states where sharp avoided crossings are likely to appear~\cite{WangPRA2011,Kokoouline2006PRA,Suno2011JChemPhys}. Future work will include extending the current derivation to the discrete variable representation (DVR) central to the SVD technique of Ref.~\cite{TolstikhinJPB1996}.
 
\section{Regularization of interactions in the two-body, multi-channel model \label{sec:regularization}}

Although the bound states and couplings to them in Sec. III have a finite range, since they are small compared to $\lho$ we find it convenient to work with the zero-range limit of these couplings, for example in Eq.~\eqref{eq:Wnb-factor}. However, fixing the bound state energies and taking the zero-range limit is problematic and leads to divergences.

\begin{figure*}
\includegraphics[width=1.3\columnwidth]{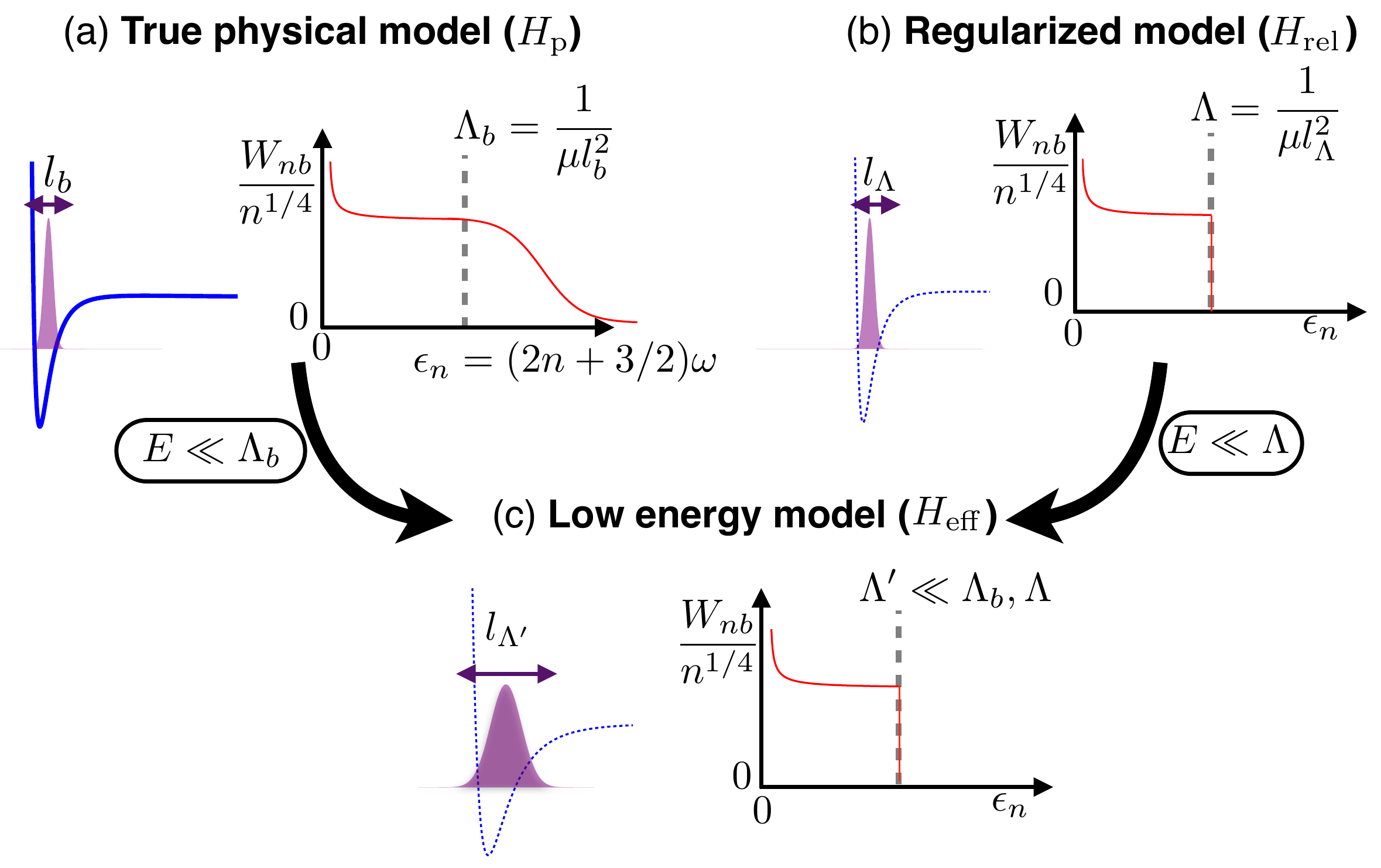}
\caption{ (Color online)
\emph{Derivation of the regularized model Eq.~\eqref{eq:H-rel-reg}.} The regularized zero-range limit (panel b) of the true physical model (panel a) is computed by ensuring the agreement of the forms of their low energy effective Hamiltonians (panel c), which govern all of the low-energy observables. Even though the true physical Hamiltonian cannot be computed, its form can be determined [see Eq.~\eqref{eq:wbexact}] and shown to match that obtained from the regularized model for appropriately chosen $W_{nb}$ and $\nu_b$.  
\label{fig:reg-cartoon}}
\end{figure*}

In this section, we describe how Eq.~\eqref{eq:H-2-body-ho}  may be regularized to obtain the physical Hamiltonian from which the effective parameters of Eq.~\eqref{eq:EffectiveModel} are calculated.  The naive zero-range limit takes the bound state energies $\nu_b$ to be fixed in the absence of a coupling $W_{nb}$, and approximates the bound state wavefunctions as having zero range.  This naive limit could be obtained from the results in Sec.~\ref{sec:Wnbfactor} by taking $l_b\to 0$, allowing us to apply Eqs.~\eqref{eq:obwbexact} and~\eqref{eq:wbexact} for all $n$.  However, the true physical limit is a bit more subtle: the \textit{physical} bound state energies are indeed some finite, fixed set of numbers, but these are not the same as the $\nu_b$ in Eq.~\eqref{eq:H-2-body-ho}. Rather, the physical energies correspond to the eigenenergies after coupling to the open channel. Analogous to the well-known one- and two-channel cases~\cite{pethick02}, this coupling to the continuum
 gives a divergent shift of the eigenenergies away from the $\nu_b$.
Although the regularization of the one- and two-channel models is standard
and requires only a (diverging) shift in the bare bound state energies, the regularization of a multi-channel model such as ours requires new forms of couplings, and to our knowledge the appropriate regularization was first presented in Ref.~\cite{docaj:ultracold_2016}. 

Figure~\ref{fig:reg-cartoon} summarizes the logic of the derivation of the regularized Hamiltonian, which we outline before giving it in detail. The basic approach is to calculate the effective low-energy Hamiltonian for the model that we claim is the proper regularized zero-range limit. Then we calculate the the effective low-energy Hamiltonian for the true multi-channel physical Hamiltonian, whose form and general properties we know, even though we do not know the values of the parameters appearing in it.  Finally, we show that parameters for the the zero-range Hamiltonian can be chosen such that its effective Hamiltonian matches the true Hamiltonian's effective Hamiltonian at low energies, thereby confirming the proposed form of the regularized Hamiltonian.

The Hamiltonian that properly accounts for the zero-range limit (as argued below) is the $\Lambda\rightarrow \infty$ limit of
\begin{widetext}
\be
\hat{H}_{\text{rel}}(\Lambda) &=& \sum_{n \text{ with } \epsilon_n<\Lambda} \epsilon_n \ket{n}\!\bra{n} +\sum_{b,b'} \lp \delta_{bb'} \nu_b^{\star} + \sqrt{\frac{\mu ^3\Lambda}{2}} w_b w_{b'}\rp \ket{b}\!\bra{b'}
+\sum_{b,n \text{ with } \epsilon_n<\Lambda} \lp \frac{w_{b} M_n}{\lho^{3/2}} \ket{n}\!\bra{b}+\hc\rp  \label{eq:H-rel-reg}
\ee
\end{widetext}
where $\Lambda$ is a high energy cutoff (short distance, $l_{\Lambda}$, cutoff) for the open channel.  This energy cutoff means that the sum over harmonic oscillator states $n$ runs only to a value $n^{\star}$ such that states with $n<n^{\star}$ have energies $\epsilon_{n}<\Lambda$.  Explicitly, $n^{\star}=\operatorname{Floor}[\Lambda/(2\omega)-3/4]$, although it suffices to take $n^{\star}=\Lambda/(2\omega)$ since we take the $\Lambda\rightarrow \infty$ limit. The key addition to Eq.~\eqref{eq:H-2-body-ho} to obtain the physical zero-range limit is the term proportional to $\sqrt{\Lambda}$ that couples bound states $\ket{b}$ and $\ket{b'}$ and shifts the energy of each bound state $\ket{b}$. 
We now show (I) that the physical properties of Eq.~\eqref{eq:H-rel-reg} are independent of $\Lambda$ for $\Lambda/\omega \gg 1$ [i.e.~Eq.~\eqref{eq:H-rel-reg} is a regularization of Eq.~\eqref{eq:H-2-body-ho}] and (II) that Eq.~\eqref{eq:H-rel-reg} can reproduce the low-energy properties of the true microscopic physical Hamiltonian [i.e.~it is the appropriate physical regularization]. 

For the first point, we want to compute $H_{\text{rel}}(\Lambda)$'s effective Hamiltonian $\hat{H}_{\text{eff}}(\Lambda')$ defined to act on the low-energy restricted Hilbert space that includes only bound states and open channel states with $\epsilon_n<\Lambda'$. The effective Hamiltonian in a restricted Hilbert space may be obtained by second-order degenerate perturbation theory (i.e.~a Schrieffer-Wolff transformation):
\be
\hat{H}_{\text{eff}}(\Lambda') &=&  \hat{H}_{\text{rel}}(\Lambda')  -\sum_{n,b,b'}\!\!{}^{{ '}} \frac{w_b w_{ b'}M_n^2}{\epsilon_n \lho^{3}} \ket{b}\! \bra{b'}
\ee  
(to leading order in $1/\Lambda'$) where  $\sum'$ indicates a sum over $n$ such that $\Lambda' < \epsilon_n < \Lambda$. The second term represents the fluctuations to the high energy Hilbert space with $\epsilon_n>\Lambda'$ that are being eliminated.  It is $-\mc J \sum_{b,b'} w_b w_{b'}/\lho^{3} \ket{b}\bra{b'}$ with $\mc J=\sum_n' M_n^2/\epsilon_n$. As $\{\Lambda,\Lambda'\}\rightarrow \infty$, ${\mc J}\approx \sqrt{\mu^3/2\lho^3}(\sqrt{\Lambda}-\sqrt{\Lambda'})$.  Adding this term to $\hat{H}_{\text{rel}}(\Lambda')$, we see that the $\sqrt{\Lambda}$ terms cancel so that $\Lambda$ in $\hat{H}_{\text{rel}}$ is effectively replaced by $\Lambda'$: 
\be
\hat{H}_{\text{eff}}(\Lambda') &=& \hat{H}_{\text{rel}}(\Lambda').\ee
This is not vacuous: it says that the effective Hamiltonian for the $\epsilon_n<\Lambda'$ subspace is simply the Hamiltonian in Eq.~\eqref{eq:H-rel-reg} -- whose $\Lambda\rightarrow \infty$ limit is used to define the theory -- evaluated at the lower cutoff $\Lambda'$. 
This shows immediately that in contrast to the unregularized Hamiltonian in Eq.~\eqref{eq:H-2-body-ho} the $\hat{H}_{\text{rel}}(\Lambda)$ is finite: since $\hat{H}_{\text{eff}}(\Lambda')$ is defined on a space with a finite cutoff and gives identical low-energy physics as $\hat{H}_{\text{rel}}$ at any value of $\Lambda$, the physics described by $\hat{H}_{\text{rel}}(\Lambda)$ is $\Lambda$-independent.  In particular, one can take $\Lambda\rightarrow\infty$ when this is convenient.
 
Now we show that not only is Eq.~\eqref{eq:H-rel-reg} finite, it reproduces the low-energy observables of the true physical Hamiltonian $H_p$. Our approach will be to compute the effective low-energy Hamiltonian at energy $\Lambda'$ associated with $\hat{H}_p$, which we denote $\hat{H}_p(\Lambda')$, and match Eq.~\eqref{eq:H-rel-reg} parameters to reproduce it. 
We can do this without knowing the details of $\hat{H}_p$: we need merely some of its general properties. In contrast to Eq.~\eqref{eq:H-rel-reg}, $W_{nb}$ factors as $w_b M_n/\lho^{3/2}$  only for sufficiently small $n$ that the harmonic oscillator wavefunction varies slowly over the length scale of the bound state. Note that this $w_b$ is associated with the physical Hamiltonian and not the regularized Hamiltonian; we avoid introducing new notation and the variables are to be distinguished by context. We will relate the $w_b$ in the two models shortly. 

At large $n$  the harmonic oscillator state probes the short-range (high-energy) physics on the scale of the bound state size and there is no simple expression for the $W_{nb}$. Nevertheless, we may formally determine the Hamiltonian describing low-energy observables below a cutoff $\Lambda'$ as above, as we know that the $W_{nb}$ fall off very rapidly at some energy scale $\Lambda_b$ (corresponding to a length scale $l_b$), see Fig.~\ref{fig:reg-cartoon}(a). Here, we choose $\omega \ll \Lambda' \ll \Lambda_b$. This guarantees that $\Lambda'$ is large enough for the trap levels to be treated as a continuum, but small enough not to probe the short range bound state structure.  Such a choice is possible since $\lho$ is much greater than the microscopic lengths characterizing the intermolecular interactions. As a consequence of this choice, for the $n$ in the truncated low-energy Hilbert space we have $W_{nb}=w_{b} M_n/\lho^{3/2} $ for some $w_b$.
To leading order in $1/\Lambda'$ we find  
\begin{widetext}
\be
\hat{H}_p(\Lambda') &=& 
\sum_{n \in {\mc L}(\Lambda')} \epsilon_n \ket{n}\!\bra{n} +\sum_{b,b'} \lp \delta_{bb'} \nu_b - \sum_{n\in {\mc H}(\Lambda')} \frac{W_{nb}W_{nb'}}{\epsilon_n} \rp \ket{b}\!\bra{b'}
+\sum_{b,n \in {\mc L}(\Lambda')} \lp \frac{w_{b}M_n}{\lho^{3/2}} \ket{n}\!\bra{b}+\hc\rp  \label{eq:Hp-eff}
\ee
\end{widetext}
where the sets ${\mc L}(\Lambda')$ and ${\mc H}(\Lambda')$ are the $n$s with $\epsilon_n<\Lambda'$ and $\epsilon_n>\Lambda'$, respectively. In the last term $W_{nb}$ is replaced with $w_b M_n/\lho^{3/2}$, which is valid because the sum is only for $\epsilon_n < \Lambda'$ and our choice of $\Lambda'$ is small enough for it to be valid.  The couplings  are all finite.

To complete our derivation, we show that the effective Hamiltonian at scale $\Lambda'$ for the physical Hamiltonian, $\hat{H}_p(\Lambda')$ can be matched by $\hat{H}_{\text{eff}}(\Lambda')$, the effective Hamiltonian associated with $\hat{H}_{\text{rel}}$.  The only apparent difference is in the middle term, the coefficient of $\ket{b}\!\bra{b'}$. To recast the two Hamiltonians in the same form, we calculate the sum $\mc S=-\sum_{n\in {\mc H}(\Lambda')} (W_{nb}W_{nb'})/\epsilon_n$ appearing in Eq.~\eqref{eq:Hp-eff}. We split the sum into two pieces $\mc S={\mc S}_1+{\mc S}_2$: (i) ${\mc S}_1$ sums in an energy range from $\Lambda'$ to $\Lambda^{\star}$, where $\Lambda^{\star}$ 
is chosen so that below $\Lambda^{\star}$ the bound state structure is not probed and the factorization $W_{nb}=w_b M_n/\lho^{3/2}$ is valid, and (ii) ${\mc S}_2$ sums from $\Lambda^{\star}$ to $\infty$. The first sum is ${\mc S}_1=-w_b w_{b'}\sqrt{\mu^3/2}(\sqrt{\Lambda^{\star}}-\sqrt{\Lambda'})$. The second term is finite, and independent of $\Lambda'$; we denote it $f_{b,b'}$. Thus we can write ${\mc S}=w_b w_{b'}\sqrt{\mu^3\Lambda'/2}+g_{b,b'}$ where $g_{b,b'}$ is finite and independent of the cutoff $\Lambda'$. With this evaluation of  ${\mc S}$, we see that the effective Hamiltonians $\hat{H}_p(\Lambda')$ and $\hat{H}_{\text{rel}}(\Lambda)$ differ only by the $g_{b,b'}$ term in the former. We can diagonalize the matrix consisting of the $\nu_b$ and $g_{b,b'}$ terms; following this basis transformation to a basis $\overline{\ket{b}}$, one obtains
\begin{widetext}
\be 
\hat{H}_p(\Lambda') &=& 
\sum_{n \in {\mc L}(\Lambda')} \epsilon_n \ket{n}\!\bra{n} +\sum_{b,b'} \lp -\delta_{bb'} {\bar \nu}_b -  \sqrt{\frac{\mu^3 \Lambda}{2}}{\bar w}_b {\bar w}_{b'}\rp \overline{\ket{b}}\!\overline{\bra{b'}}
+\sum_{b,n \in {\mc L}(\Lambda')} \lp \frac{{\bar w}_{b}M_n}{\lho^{3/2}} \ket{n}\!\bra{b}+\hc\rp  \label{eq:Hp-cutoff-explicit}
\ee
\end{widetext}
where ${\bar w}_b$ and ${\bar\nu}_b$ are the couplings following the basis transformation.  Note that we have chosen the factors involved in the diagonalization such that the transformed basis vectors and couplings do not depend on the cutoff. Identifying the $\overline{\ket{b}}$, ${\bar w}_b$, and ${\bar \nu}_b$ in Eq.~\eqref{eq:Hp-cutoff-explicit} with the $\ket{b}$, $w_b$, and $\nu^{\star}_b$ in Eq.~\eqref{eq:H-rel-reg} we see that $\hat{H}_p(\Lambda')$ exactly coincides with $\hat{H}_{\text{rel}}(\Lambda')$. 

To summarize, we have shown that we are able  to use an effective  model that ignores the true structure of the couplings $W_{nb}$ at high energy and assumes that the zero-range bound state approximation is valid to determine the couplings $W_{nb}=w_b M_n/\lho^{3/2}$  for all $n$. The cost is to add diverging bound state-bound state couplings.  Upon doing this, our Hamiltonian exactly reproduces the low energy observables of the true physical Hamiltonian.

\section{Derivation of lattice model \label{sec:latt-mod-derive}}

In this section, we derive the multi-channel Hubbard model Eq.~\eqref{eq:EffectiveModel} starting from a microscopic continuum description.  Namely, the continuum description of our model may be written as
\begin{align}
\hat{H}&=\hat{H}_{\mathrm{internal}}+\hat{H}_{\mathrm{kin}}+\hat{H}_{\mathrm{latt}}+\hat{H}_{\mathrm{couple}}\, .
\end{align}
Here, $\hat{H}_{\mathrm{internal}}$ is the Hamiltonian of the internal energy of the molecules and BCCs, which can be written in terms of the molecular annihilation field operators $\hat{\psi}_s\left(\mathbf{r}\right)$ and the BCC annihilation field operators $\hat{\psi}_{b;\mathrm{BCC}}\left(\mathbf{r}\right)$ as
\begin{align}
\nonumber \hat{H}_{\mathrm{internal}}&= \sum_b\int \!d\mathbf{r}\,E_b\hat{\psi}^{\dagger}_{b;\mathrm{BCC}}\left(\mathbf{r}\right)\hat{\psi}_{b;\mathrm{BCC}}\left(\mathbf{r}\right)\\
&+\sum_s\int \!d\mathbf{r}\, E_s\hat{\psi}^{\dagger}_{s}\left(\mathbf{r}\right)\hat{\psi}_{s}\left(\mathbf{r}\right) \, ,
\end{align}
where $E_b$ is the energy of BCC state $|b\rangle$ and $E_s$ is the energy of open channel internal state $|s\rangle$.  The next term
\begin{align}
\nonumber \hat{H}_{\mathrm{kin}}=&-\sum_{s}\int \!d\mathbf{r}\,\hat{\psi}_s^{\dagger}\left(\mathbf{r}\right)\frac{1}{2m}\nabla^2\hat{\psi}_s\left(\mathbf{r}\right)\\
&-\sum_b\int \!d\mathbf{r}\,\hat{\psi}_{b;\mathrm{BCC}}^{\dagger}\left(\mathbf{r}\right)\frac{1}{4m}\nabla^2\hat{\psi}_{b;\mathrm{BCC}}\left(\mathbf{r}\right)\, ,
\end{align}
with $m$ the molecular mass, is the kinetic energy operator.  The next term is the lattice potential energy, given by
\begin{align}
&\label{eq:Hlatt}\hat{H}_{\mathrm{latt}}=\sum_s\int \!d\mathbf{r}\,\hat{\psi}_s^{\dagger}\left(\mathbf{r}\right) V\left(\mathbf{r}\right)\hat{\psi}_s\left(\mathbf{r}\right)\\
\nonumber &+\sum_b \int \!d\mathbf{r}\,\hat{\psi}_{b;\mathrm{BCC}}^{\dagger}\left(\mathbf{r}\right) 2V\left(\mathbf{r}\right) \hat{\psi}_{b;\mathrm{BCC}}\left(\mathbf{r}\right)\, ,
\end{align}
with $V\left(\mathbf{r}\right)$ a periodic lattice potential.  For simplicity, throughout this section we assume that the lattice is ``magic" in the sense that the depth is independent of the internal state of the NRMs.  This is an excellent approximation for hyperfine states, and can be achieved for rotational states through polarization~\cite{PhysRevLett.109.230403} or electric field~\cite{PhysRevA.82.063421} control.  Such magic lattices are desirable for reducing dephasing due to spatially inhomogeneous light shifts.  Our model can be straightforwardly generalized to non-magic conditions.  We also take the polarizability of the BCCs equal to twice the molecular polarizability; the consequences of relaxing this condition are explored in Ref.~\cite{wall:beyond_2016}.  Finally, the last term in the Hamiltonian,
\begin{align}
\hat{H}_{\mathrm{couple}}&=\sum_{b,s,s'}\int \!d\mathbf{r}\,\left[\hat{\psi}^{\dagger}_{b;\mathrm{BCC}}\left(\mathbf{r}\right)W_{s,s',b} \hat{\psi}_s\left(\mathbf{r}\right) \hat{\psi}_{s'}\left(\mathbf{r}\right)+\mathrm{H.c.}\right]\, ,
\end{align}
is the pairing of two NRMs via a short-range coupling with matrix elements ${W}_{s,s',b}$ to form a BCC.  Similar continuum resonance models have been used to model Feshbach resonances for ultracold atoms in optical lattices~\cite{Timmermans1999199,PhysRevLett.95.243202,Gurarie20072,PhysRevLett.104.090402,von_Stecher_Gurarie_11,PhysRevLett.109.055302,PhysRevA.87.033601}.

To derive a lattice model from the continuum description above, we expand the field operators in a complete set of localized Wannier functions; we denote the Wannier functions of the molecules as $w_{i,n,s}\left(\mathbf{r}\right)$ and those of the BCCs as $\mathcal{W}_{i, n_b,b}\left(\mathbf{r}\right)$, where $i$ is the site index and $n$ and $n_b$ are band indices.  Explicitly, these expansions read
\begin{align}
\hat{\psi}_s\left(\mathbf{r}\right)&=\sum_{i,n}w_{i,n,s}\left(\mathbf{r}\right) \hat{a}_{i,n,s}\,, \\
\hat{\psi}_{b;\mathrm{BCC}}\left(\mathbf{r}\right)&=\sum_{i,n_b}\mathcal{W}_{i,n_b,b}\left(\mathbf{r}\right) \hat{\mathcal{A}}_{i,n_b,b}\, ,
\end{align}
where $\hat{a}_{i,n,s}$ and $\hat{\mathcal{A}}_{i,n_b,b}$ are annihilation operators acting on the associated Fock spaces for molecules and BCCs, respectively.  

With these definitions, and under conditions analogous to those required for the validity of the Hubbard model description for atoms -- specifically that (i) tunneling can be truncated to nearest neighbors, (ii) all molecules not in short-range bound states are in the lowest band of the lattice, (iii) tunneling of the BCCs can be neglected, and (iv) pairing processes (i.e.~matrix elements of $\hat{H}_{\mathrm{couple}}$) occurring between molecules on different lattice sites can be neglected -- we find the multi-channel Hubbard model
\begin{align}
\label{eq:Hred}\hat{H}&=\hat{H}_{\mathrm{on-site}}-J\sum_{\langle i,j\rangle,s}\left[\hat{a}_{i,s}^{\dagger}\hat{a}_{j,s}+\mathrm{H.c.}\right]\, ,
\end{align}
where $\hat{a}_{i,s}\equiv \hat{a}_{i,0,s}$ and $\langle i,j\rangle$ denotes a sum over nearest-neighbor pairs $i$ and $j$.  Justifications for the conditions (i)-(iv) will be given shortly.  In Eq.~\eqref{eq:Hred}, the on-site Hamiltonian is
\begin{align}
&\hat{H}_{\mathrm{on-site}}=\sum_{i,n,s} E_{n,s}\hat{n}_{i,n,s}+\sum_{i,n_b,b}\mathcal{E}_{n_b,b}\hat{\mathcal{N}}_{i,n_b,b}\\
\nonumber &+\sum_{i,s,s',b,n,n',n_b}\left[W_{s,s',b}^{n,n',n_b}\hat{\mathcal{A}}_{i,n_b,b}^{\dagger}\hat{a}_{i,n,s}\hat{a}_{i,n',s'}+\mathrm{H.c.}\right]\, ,
\end{align}
with molecule energies $E_{n,s}=\varepsilon_n+E_s$, $\varepsilon_n$ the average of the band structure for band $n$ over the first Brillouin zone, BCC energies $\mathcal{E}_{n_b,b}=E_b+\varepsilon_{n_b}$, $\hat{n}_{i,n,s}=\hat{a}^{\dagger}_{i,n,s}\hat{a}_{i,n,s}$, $\hat{\mathcal{N}}_{i,n_b,b}=\hat{\mathcal{A}}^{\dagger}_{i,n_b,b}\hat{\mathcal{A}}_{i,n_b,b}$, and we have defined the overlap integrals
\begin{align}
 &J=-\int d\mathbf{r}w_{i,0,s}\left(\mathbf{r}\right)\Big[-\frac{1}{2m} \nabla^2+V\left(\mathbf{r}\right)\Big]w_{j,0,s}\left(\mathbf{r}\right)\, ,\\
&W_{s,s',b}^{n,n',n_b}={W}_{s,s',b}\int d\mathbf{r}\mathcal{W}_{i,n_b,b}\left(\mathbf{r}\right)w_{i,n,s}\left(\mathbf{r}\right)w_{i,n',s'}\left(\mathbf{r}\right) \, ,
\end{align}
in which $w_{j,0,s}\left(\mathbf{r}\right)$ is a nearest-neighbor of $w_{i,0,s}\left(\mathbf{r}\right)$.  

We now turn to the justification of the conditions (i)-(iv) above.  Condition (i) is standard for atomic Hubbard models, and follows from the exponential decrease of tunneling amplitude with distance tunneled.  Condition (ii) is reasonable for near-term experiments, in which molecules are created from ultracold atomic gases in the lowest band and transfer of population to higher bands during molecule formation can be ignored~\cite{PhysRevA.92.063416,covey2016doublon}.  Also, while the complex dynamics of $\hat{H}_{\mathrm{on-site}}$ can involve mixing between bands, such configurations only exist on the scale of a single lattice site.  Condition (ii) only requires that NRMs in higher bands occur only in such short-range bound configurations, and are not free to propagate through the lattice.

To illustrate the validity of (iii) and (iv), we will specialize to the case of a simple cubic lattice, for which $V\left(\mathbf{r}\right)=V\sum_{\nu=x,y,z}\sin^2(\pi \nu/a)$ with $a$ the lattice spacing.  Because of the additive separability of this lattice potential, tunneling occurs only along the principal axes.  The nearest-neighbor tunneling amplitude along a principal axis, determined by a fit to numerically generated data in the range $V/E_R\ge 2$, is
\begin{align}
\frac{J}{E_R}&\approx 1.363 \left(\frac{V}{E_R}\right)^{1.057}\exp\left(-2.117\sqrt{V/E_R}\right)\, ,
\end{align}
where $E_R=\pi^2/2ma^2$ is the recoil energy.  For the BCCs, the recoil energy is half that of a molecule, and the depth roughly twice, and so the ratio $V/E_R$ is four times larger than for molecules.  In light of the exponential dependence of tunneling on lattice depth, this makes the tunneling of BCCs negligible compared to the tunneling of molecules.  For example, the tunneling of BCCs is $\sim 1\%$ of the molecular tunneling for a molecular lattice depth of $V\sim 8E_R$.  However, we stress that although the tunneling rate of BCCs is significantly smaller than that of molecules, BCCs and molecules have the same harmonic oscillator trapping frequency, as the increase in the lattice depth and mass cancel in $ \omega=2\sqrt{V E_R}$, within the approximation that the polarizability of BCCs is twice that of molecules.

We now turn to the matrix elements of $\hat{H}_{\mathrm{couple}}$.  These separate into a coupling constant ${W}_{s,s',b}/a^{3/2}\sim w_b/a^{3/2}$ with units of energy and a dimensionless geometric integral.  For on-site coupling, this dimensionless integral is $I_{n_b;n,n'}=a^{3/2}\int d\mathbf{r}\mathcal{W}_{i,n_b,b}\left(\mathbf{r}\right)w_{i,n,s}\left(\mathbf{r}\right)w_{i,n',s'}\left(\mathbf{r}\right)$.  In the approximation where each site is a harmonic well, which becomes exact as $V\to \infty$, this geometric integral has the scaling $\sim \lho^{-3/2}=(V/E_R)^{3/8}$; a fit to numerical data in the range $V/E_R\in\left[12,45\right]$ yields $I_{0;0,0}\sim (V/E_R)^{0.42}$, in reasonable agreement.  For the off-site pairing term, the most relevant geometric integral is $\tilde{I}=a^{3/2}\int d\mathbf{r}\mathcal{W}_{i,0,b}\left(\mathbf{r}\right)w_{i,0,s}\left(\mathbf{r}\right)w_{i+1,0,s'}\left(\mathbf{r}\right)$, in which $w_{i+1,0,s'}\left(\mathbf{r}\right)$ is shifted from $w_{i,0,s}\left(\mathbf{r}\right)$ by a single lattice spacing along one principal axis.  We find numerically that the scaling of this integral is $\tilde{I}\sim 0.36 (\frac{V}{E_R})^{0.16}\frac{J}{E_R}$, and so is much smaller than the on-site coupling.  In addition to the geometric integral being small, the actual magnitude of this process is small even compared to tunneling because of the disparity in energy scales, $(w_b/a^{3/2})/E_R\ll 1$.  Hence, in contrast to resonance models for broad Feshbach resonances, in which off-site pairing of molecules is a key process~\cite{PhysRevLett.95.243202,von_Stecher_Gurarie_11,PhysRevLett.109.055302}, due to the narrowness of the resonances experienced by NRMs (set by the high density of BCCs at zero energy~\cite{mayle:statistical_2012,mayle:scattering_2013,docaj:ultracold_2016,wall:beyond_2016}), such processes are irrelevant and can be safely ignored.  

We next turn to the solution of $\hat{H}_{\mathrm{on-site}}$, which encapsulates the complex short-range physics that occurs on the length scale of a single site.  For simplicity, we solve $\hat{H}_{\mathrm{on-site}}$ in the harmonic oscillator approximation for a single lattice site, as the separation of the center of mass and relative coordinates reduces the effective dimensionality of the problem.  We further assume that there are never more than two molecules on a given lattice site, which occurs in many contexts, e.g., when one works close to an $n=1$ Mott insulator (even in the adjacent strongly-interacting superfluid phase), or in optical microtraps~\cite{Kaufman2012,PhysRevLett.110.133001,hutzler2016eliminating} where the number of particles can be precisely monitored.  The zero-NRM and (relevant) one-NRM sectors of $\hat{H}_{\mathrm{on-site}}$ on site $i$ are trivially spanned by the vacuum $|\mathrm{0}\rangle_i$ and $\hat{a}^{\dagger}_{i,0,s}|\mathrm{0}\rangle_i=|s\rangle_i$, respectively.  For two molecules, the Hamiltonian can be separated into center of mass and relative coordinates, and yields $\hat{H}_{{\text{c.m.}}}$ and $\hat{H}_{\mathrm{rel}}$ given in Eq.~\eqref{eq:H-2-body-ho} and the surrounding discussion.  We note that the pairing in the relative coordinate Hamiltonian depends on the internal state configuration of the open channel NRMs indexed by $s$ and $s'$ and the zero of energy is taken to be $E_s+E_{s'}$.  The regularization of this on-site Hamiltonian was discussed in detail in Sec.~\ref{sec:regularization}.  Here, we only need to note that the two-particle relative coordinate solutions have the general form ($s$-wave coupling assumed)
\begin{align}
|\alpha\rangle_i&=\sum_{n,s,s'}p_{\alpha;n,s,s'}|n,\ell=0\rangle|s,s'\rangle_i+\sum_{b}q_{\alpha;b}|b\rangle_i\, ,
\end{align}
and energies $E_{\alpha}$.  

With the two-body solution in hand, we can derive an effective model which is valid at low on-site density by considering only on-site configurations in which there is no molecule (vacuum state $|0\rangle_i$), a single molecule in one of the open-channel states $\hat{a}^{\dagger}_{i,s}|0\rangle=|s\rangle_i$, or one of the two-molecule eigenstates $|\alpha\rangle_i$ which is near-resonant with two separated molecules on the scale of the coupling.  We do so by noting that we can re-express doubly occupied sites as
\begin{align}
\hat{a}^{\dagger}_{i,s'}|s\rangle_i&=P_{s,s'}\sqrt{1+\delta_{s,s'}}|s,s'\rangle_i\, ,\\
&=\sum_{\alpha} P_{s,s'}\sqrt{1+\delta_{s,s'}}_i\langle \alpha|s,s'\rangle_i |\alpha\rangle_i\, ,
\end{align}
by using the completeness of two-molecule eigenstates $|\alpha\rangle_i$\footnote{This relation also assumes that the matrix element connecting the relevant spherical harmonic oscillator quantum numbers and band quantum numbers (Talmi-Moshinsky coefficient) is unity.  This is the case for the $|n_{\mathrm{c.m.}}=0\rangle$ center-of-mass state, $|n=0\rangle$ relative coordinate state, and the lowest band; for other motional configurations and additional overlap integral is required}.  Here, the square-root factor, which only contributes for identical bosons, accounts for Bose stimulation, we can identify ${}_i\langle \alpha|s,s'\rangle_i\equiv p_{\alpha;0,s,s;}=\mathcal{O}_{\alpha}^{s,s'}$ as the opacity, and $P_{s,s'}$ accounts for on-site fermionic exchange and Pauli blocking.  It is 1 for bosons, and
\begin{align}
P_{s,s'}&=\left\{\begin{array}{c} 0\;\;\; s=s'\\ 1\;\;\; s> s'\\ -1\;\;\; s<s'\end{array}\right.\, ,
\end{align}
for fermions.  Hence, we can concisely capture the low-filling constraint by defining modified operators as
\begin{align}
\label{eq:Modops}\hat{c}^{\dagger}_{i,s}&=\hat{b}^{\dagger}_{i,s}+\sum_{\alpha}\sum_{s'}\sqrt{1+\delta_{s,s'}}P_{ss'}\mathcal{O}_{\alpha}^{s,s'}\hat{d}^{\dagger}_{i,\alpha}\hat{b}_{i,s'}\,,
\end{align}
where $\hat{d}^{\dagger}_{i,\alpha}$ is a hard-core bosonic operator which creates an NRM pair in the eigenstate $|\alpha\rangle_i$ and $\hat{b}^{\dagger}_{i,s}=\hat{\mathcal{P}}^{(1)}\hat{a}^{\dagger}_{i,s}\hat{\mathcal{P}}^{(0)}$ with $\hat{\mathcal{P}}^{(n)}$ a projector onto the subspace with exactly $n$ NRMs on a site.  On different sites the operators $\hat{c}_{i,s}$ are defined to have the same (anti)commutation relations as the ``bare" operators $\hat{a}_{i,s}$.  In can be verified that this construction is the same as Eqs.~\eqref{eq:c1}-\eqref{eq:c3} above.  In terms of these operators, the tunneling term in Eq.~\eqref{eq:Hred} with the low filling constraint becomes
\begin{align}
-J\sum_{\langle i,j\rangle,s}\left[\hat{a}_{i,s}^{\dagger}\hat{a}_{j,s}+\mathrm{H.c.}\right]&\to -J\sum_{\langle i,j\rangle, s}\left[ \hat{c}^{\dagger}_{i,s}\hat{c}_{j,s}+\mathrm{H.c.}\right]\, .
\end{align}
The on-site Hamiltonian $\hat{H}_{\mathrm{on-site}}$ can be similarly transformed to the low-filling subspace by defining number operators $\hat{n}_{i\alpha}=|\alpha\rangle_i\langle \alpha|_i$ and $\hat{n}_i=\sum_{s}|s\rangle_i\langle s|_i+2\sum_{\alpha}|\alpha\rangle_i\langle \alpha|_i$ to write
\begin{align}
\hat{H}_{\mathrm{on-site}}&\to \sum_i\left(\sum_{\alpha} U_{\alpha} \hat{n}_{i,\alpha}+\frac{3\omega}{2}\hat{n}_i\right)\, ,
\end{align}
where we have partitioned the energy of a two-molecule eigenstate into interaction and trap components as $E_{\alpha}=U_{\alpha}+3\omega/2$ by defining $U_{\alpha}=E_{\alpha}-3\omega/2$.  Putting the tunneling and on-site terms together, we arrive at
\begin{align}
\hat{H}=&-J\sum_{\langle i,j\rangle, s}\left[ \hat{c}^{\dagger}_{i,s}\hat{c}_{j,s}+\mathrm{H.c.}\right]+\sum_i\left(\sum_{\alpha} U_{\alpha} \hat{n}_{i,\alpha}+\frac{3\omega}{2}\hat{n}_i\right)\, ,
\end{align}
which is Eq.~\eqref{eq:EffectiveModel}.

To give some physical insight into the processes occurring in this multi-channel resonance model (Fig.~\ref{fig:latt-eff-model}), it is useful to replace the definitions of the low-filling operators into Eq.~\eqref{eq:EffectiveModel} to find 
\begin{widetext}
\begin{align}
\label{eq:MCEM} \hat{H}=&\sum_i\left[\sum_{\alpha} U_{\alpha}\hat{d}^{\dagger}_{i\alpha}\hat{d}_{i\alpha}+\frac{3}{2}\omega\left(\sum_{s}\hat{b}_{i,s}^{\dagger}\hat{b}_{i,s}+2\sum_{\alpha} \hat{d}^{\dagger}_{i\alpha}\hat{d}_{i\alpha}\right)\right]\\
\nonumber&-J\sum_{\langle i,j\rangle}\sum_{s} \left[\left(\hat{b}_{i,s}^{\dagger}+\sum_{\alpha,s'}\sqrt{1+\delta_{s,s'}}P_{ss'}\mathcal{O}_{\alpha}^{s,s'}  \hat{d}^{\dagger}_{i\alpha}\hat{b}_{i,s'}\right)\left(\hat{b}_{j,s}+\sum_{\alpha',s''}\sqrt{1+\delta_{s,s''}}P_{ss''}\mathcal{O}_{\alpha'}^{s,s''}\hat{d}_{j\alpha'}\hat{b}^{\dagger}_{j,s''}\right)+\mathrm{H.c.}\right]\, .
\end{align}
\end{widetext}
The first line is the on-site energy, which correctly reproduces the $3\omega/2$ trap energy of each molecule and the additional $U_{\alpha}$ interaction energy when two NRMs share a lattice site.  The second line consists of three different tunneling processes which can be organized according to powers of the opacities $\mathcal{O}_{\alpha}^{s,s'}$.  The first process is tunneling of an NRM from a singly-occupied site to an unoccupied site, given by the term in the second line with only $\hat{b}$ operators.  This process occurs at the bare tunneling rate $J$.  The next-order process is tunneling of a molecule from singly occupied site to another site which also contains a single molecule (and its Hermitian conjugate).  This process ends with two molecules on a single site, and so the resulting state is projected onto the two-molecule eigenstates $|\alpha\rangle_i$, yielding a single factor of the opacity $\mathcal{O}_{\alpha}^{s,s'}$.  Hence, the tunneling of NRMs onto occupied sites occurs at the slower rates $\mathcal{O}_{\alpha}^{s,s'}J$.  The final process is when a doubly occupied site and a neighboring singly occupied site exchange positions, and is described by the product of the terms involving $\hat{d}$ operators in the second line.  Here, operationally, the doubly occupied site is projected from the two-body eigenstates $|\alpha\rangle_i$ into open channel states $|s,s'\rangle$, giving one factor of the opacity $\mathcal{O}_{\alpha}^{s,s'}$, one of the open-channel NRMs tunnels to join the neighboring site in state $|s''\rangle$, and then this two-molecule state is projected onto eigenstates $|\alpha'\rangle_i$, yielding an additional factor of the opacity $\mathcal{O}_{\alpha'}^{s,s''}$.  This ``correlated exchange" occurs at the rate $\mathcal{O}_{\alpha}^{s,s'}\mathcal{O}_{\alpha'}^{s,s''}J$, which is typically much slower than either of the other two tunneling processes.  We note that no tunneling occurs between neighboring sites when both contain two molecules due to our low-filling constraint.

\section{Conclusions and outlook \label{sec:outlook}}

We have derived an effective multi-channel Hubbard model describing ultracold nonreactive molecules (NRMs) in an optical lattice, starting from a fully microscopic description of two interacting NRMs in terms of their four constituent atoms.  Namely, from the formal four-atom description, we derived a multi-channel model for two NRMs in a harmonic trap, and discussed how to properly regularize this model to remove divergences associated with zero-range closed channel couplings.  From the solutions of this two-body model, we then derived an effective many-body lattice model under the constraints of no more than two molecules per lattice site by coupling the long-wavelength physics of NRMs in the lowest band with the two-NRM, on-site description.  In addition, we generalized the effective model beyond the description in Ref.~\cite{docaj:ultracold_2016} to include multiple internal states, such as hyperfine, rotational, or vibrational excitations, which are required to describe emerging experiments with fermionic molecules.   Our work shows that the form of the multi-channel Hubbard model Eq.~\eqref{eq:EffectiveModel} is more general than the approximations used in Ref.~\cite{docaj:ultracold_2016} to derive it. We also note that while the present focus has been on NRMs, we expect that a similar microscopic analysis, and hence the same effective model, holds for other systems which display a large density of resonant states at low scattering energy, such as have recently been observed~\cite{PhysRevA.89.020701,frisch2014quantum,maier2015broad,maier2015emergence} in highly magnetic atoms~\cite{PhysRevLett.104.063001,PhysRevLett.107.190401,1367-2630-17-4-045006,PhysRevA.85.051401,PhysRevLett.108.210401}.

The present work clearly identifies that the \emph{form} of the effective model Eq.~\eqref{eq:EffectiveModel} may be considered as exact although with unknown parameters, e.g., the interaction energies $U_{\alpha}$, and provides a formal means to determine these parameters from a fully microscopic description.  However, determining the effective model parameters by this means is a daunting task, and so other, approximate methods are desired.  One such framework for obtaining statistical distributions for the effective model parameters based on combining random matrix theory, quantum defect theory, and transition state theory, was presented in Ref.~\cite{docaj:ultracold_2016}, building on earlier ideas from Refs.~\cite{mayle:statistical_2012,mayle:scattering_2013}.  While this framework is expected to capture the qualitative structure of the model parameters, their quantitative values and the consequences of their breakdown are less certain.  In future work, it will be interesting to study this framework critically, in order to provide a computationally tractable means for obtaining effective model parameters and setting expectations for NRM experiments.  Ref.~\cite{wall:beyond_2016} begins this examination.  In addition, the breakdown of one or more of these approximations can provide insight into the general validity of these ubiquitous approximations in cold collisions and chemical physics.

In addition, while the present work defines the appropriate effective model, we have not investigated its many-body properties in any detail.  We expect that the multi-channel interaction of Eq.~\eqref{eq:EffectiveModel} can lead to significant qualitative differences in many-body physics compared to the ordinary single-channel Hubbard model.  Future many-body calculations of the equilibrium and non-equilibrium physics of Eq.~\eqref{eq:EffectiveModel} in reasonable parameter regimes may lead to new many-body phenomena which are not present in systems without complex collisions.

\section{Acknowledgements}

We thank Jose D'Incao, Kevin Ewart, Paul Julienne, and Brandon Ruzic
for useful discussions.  KRAH acknowledges the Aspen Center for Physics, which is supported by National Science Foundation grant PHY-1066293, for its hospitality while part of this work was performed. This work was supported with funds from the Welch foundation, Grant No. C-1872.  MLW acknowledges support from the NRC postdoctoral fellowship program.

\bibliography{NRM-1site-v7}

\end{document}